\documentclass[11pt]{article}

\usepackage[
    a4paper,
    left=2.3cm,
    right=2.3cm,
    top=2.4cm,
    bottom=2.4cm
]{geometry}

\usepackage{graphicx}%
\usepackage{multirow}%
\usepackage{amsmath,amssymb,amsfonts}%
\usepackage{amsthm}%
\usepackage{mathrsfs}%
\usepackage{xcolor}%
\usepackage{textcomp}%
\usepackage{booktabs}%
\usepackage{algorithm}%
\usepackage{algorithmicx}%
\usepackage{algpseudocode}%
\usepackage{listings}%
\usepackage{bm}

\usepackage[numbers,sort&compress]{natbib}
\usepackage[
    colorlinks=true,
    linkcolor=blue,
    citecolor=blue,
    urlcolor=blue
]{hyperref}
\usepackage[nameinlink,capitalise]{cleveref}

\newcommand{\JP}[1]{{\color{black}  #1}}

\title{\JP{Learning dynamically consistent flow reconstructions from limited observations}}

\author{%
Lu Zhu$^{1}$ \quad Jacob Page$^{1,*}$\\[0.6em]
\small $^{1}$School of Mathematics and Maxwell Institute for Mathematical Sciences,\\
\small University of Edinburgh, Edinburgh EH9 3FD, UK\\[0.3em]
\small $^{*}$Corresponding author: Jacob Page: \href{mailto:jacob.page@ed.ac.uk}{jacob.page@ed.ac.uk}
}

\date{}

\begin{document}

\maketitle
\begin{abstract}
A core inverse problem in the experimental sciences is the inference of a hidden dynamical state from sparse or indirect measurements. 
There is a natural opportunity for deep learning methods here, but machine-learnt reconstruction methods typically require full state data for training. 
We present Trajectory-Consistent Network Training (TraCTra), a label-free framework that trains reconstruction networks using only partial observation sequences and a differentiable forward model. 
TraCTra requires the network reconstruction and dynamical evolution to be mutually consistent: network-predicted states are marched forward in time to match subsequent observations and to agree in the full state space with independent reconstructions at later times. 
Across four fluid systems, the same objective reconstructs three-dimensional turbulence from coarse-grained fields, velocity from observations of density fluctuations, and three-dimensional density and velocity from sequences of projected two-dimensional shadowgraphs, while also recovering global vorticity from observations confined to a small spatial window. 
TraCTra outperforms assimilation-only and physics-informed neural approaches, preserves dynamically important multiscale structure, and remains accurate beyond the optimisation window. 
It transfers to held-out times in the three-dimensional shadowgraph problem and, when trained across trajectories, generalises to unseen flows in the two-dimensional problem. 
The results establish trajectory consistency as a general supervision principle for reconstructing hidden dynamical states without full state training targets.
\end{abstract}

\vspace{1em}

\section{Introduction}\label{sec:intro}

Practical measurements of turbulent flows rarely provide complete, time-resolved, three-dimensional information about the underlying flow field. Techniques such as PIV~\citep{adrian1991particle}, shadowgraphy \citep{settles2001schlieren}, and sparse sensor measurements are often limited by optical access, field of view, spatial/temporal resolution, and noise, leading to sparse, gappy, low-resolution, or indirect observations \citep{callaham2019robust}. These limitations can obscure coherent structures, energy-transfer pathways, mixing processes, and transport mechanisms. Recent advances in machine learning offer a promising route to reconstruct missing flow information from incomplete observations by learning spatial and temporal correlations from available data \citep{brunton2020machine,callaham2019robust}.

Many existing machine-learning approaches require full-field reference data during training, typically in the form of complete flow snapshots used as supervised reconstruction targets.
\JP{
For example, convolutional neural networks have been used for super-resolution reconstruction of turbulence from coarse inputs \citep{fukami2019super}, attention-based architectures like the ``Senseiver’’ \citep{santos2023development} have reconstructed complex spatial fields from coarse inputs, and related sensor-to-field approaches have mapped irregular measurements onto structured grids to form suitable inputs for convolutional processing \citep{fukami2021global}. 
}
\JP{
Other studies have cast missing-flow reconstruction as image inpainting using deep generative models \citep{buzzicotti2021reconstruction}, or have reconstructed flows through learned low-dimensional latent representations combined with regression from limited measurements \citep{dubois2022machine}.
}
These studies demonstrate the strong potential of machine learning for flow reconstruction, although they typically rely on the existence of a large library of full-field training targets. 
\JP{This can be a major practical limitation} since acquiring complete spatial fields, even over a limited interval, is often experimentally difficult or prohibitively expensive.

One possible way to deal with this limitation is to incorporate additional prior information about the system \JP{into the learning problem. }
In this context, physics-informed neural networks (PINNs)\citep{raissi2019physics} have attracted substantial attention as they \JP{incorporate governing equations, boundary conditions or other known constraints into the loss function},
\JP{and have been successfully applied in} a wide range of problems in fluid mechanics \citep{karniadakis2021physics,cai2021flow,zhu2024new}.
The physics-based losses provide extra information about unobserved or latent variables and can therefore improve reconstruction from sparse or incomplete data. 
For example, \JP{PINN-based methods} have inferred velocity and pressure fields from passive-scalar flow visualizations \citep{raissi2020hidden}, reconstructed missing pressure information and augmented experimental velocity/density measurements in stratified flows \citep{zhu2024new}, been used for sparse or incomplete flow-field reconstruction \citep{xu2025preprocessing} and for physics-informed super-resolution and denoising of fluid flows \citep{lai2026vorticity}.
\JP{However, conventional PINNs are often implemented as coordinate-based networks that map space-time locations to physical variables, which can make it difficult to exploit architectures designed for multiscale gridded data, such as convolutional, recurrent, or attention-based models. More generally, PINN training can suffer from loss imbalance, poor convergence, and gradient pathologies in nonlinear, high-dimensional, or multiscale systems~\citep{wang2021understanding,krishnapriyan2021characterizing}. These challenges motivate alternative ways of imposing physical consistency while retaining the flexibility of modern neural state-estimation architectures.}

\JP{
The machine-learning approaches described above complement classical data assimilation algorithms for state estimation from limited observations \citep[for a recent review in the context of turbulent fluid flows, see][]{ZakiARFM2025}. 
In four-dimensional variation data assimilation, or 4DVar, state estimation is framed as an optimisation problem: 
A high-resolution initial condition is sought which, when marched forward in time and observed, most closely matches the time series of observations stored from the reference experiment (for example); see \citep{Talagrand_Courtier_1987,Li_Zhang_Dong_Abdullah_2020,wang2021state}. 
This formulation does not require full-field reference states and is therefore attractive for experimental reconstruction problems in which only partial, low-resolution, or indirect measurements are available. 
The 4DVar algorithm has the benefit of being a per-trajectory optimisation — in contrast to the super-resolution approaches above which make predictions based on single under-resolved snapshots — but suffers from know deficiencies in accurate reproduction of small-scales \cite{wang2021understanding}.  
}

\JP{
Recent work has begun to combine these ideas by using 4DVar style objectives to train neural networks for flow reconstruction.
\citet{page2025super} used this approach to perform super-resolution in two-dimensional turbulence, while \citet{weyrauch2026state} adapted the ideas to three-dimensional homogeneous and isotropic turbulence (note also the study from \citet{Cleary_etal_2025} which takes the opposite approach and performs 4DVar with observation operator being defined by a pre-trained deep encoder). 
These studies showed that variational, observation-space losses can train neural reconstructors without paired high-resolution reference data. 
However, the training signal in these approaches remains essentially assimilation-like: a reconstructed state is advanced with a model and penalised according to its mismatch with later observations after application of the observation operator. 
This can be effective when the observations are direct coarse versions of the desired state, but it becomes insufficient for more ambiguous settings involving indirect measurements, derived quantities, or spatially localised observation windows. 
In such cases, matching observations alone does not ensure that the inferred high-dimensional state is dynamically consistent with the state that would be reconstructed from later measurements.
Other efforts to avoid the need for fully resolved reference data have taken a PINN-like \cite{karniadakis2021physics} approach, constraining the output fields to satisfy the Navier-Stokes equations on the coarse observation scale \citep{kelshaw2022physics}.
}

\JP{
Here we introduce Trajectory-Consistent Network Training, or TraCTra, a training framework that combines classical 4DVar observation-space consistency with an additional state-space trajectory-consistency constraint. 
In TraCTra, a network reconstruction from an initial observation is advanced with a differentiable flow solver and compared not only with later observations, but also with a separate reconstruction obtained by applying the same network to those later observations. 
The method therefore enforces consistency between two routes to the same future state: reconstructing first and then evolving the state, or evolving the physical system and reconstructing later. 
This state-space consistency converts a time series of partial observations into a high-dimensional self-supervised training signal, while avoiding the need for full-state training labels.
In the remainder of this paper we introduce the TraCTra algorithm (\S \ref{sec:workflow}) before applying it to four problems of increasing difficulty in \S \ref{sec:state_est}: super-resolution in three-dimensional turbulence, cross-modal field reconstruction, full-field reconstruction from two-dimensional shadowgraph measurements, and reconstruction of the full flow from observations restricted to a small spatial window. 
Discussion and conclusions are provided in \S \ref{sec:conclusion}.
}

\section{\JP{Trajectory-consistent network training}}\label{sec:workflow}
\subsection{\JP{Assimilation-based loss functions}}

\begin{figure}
\centering
\includegraphics[width=0.99\textwidth]{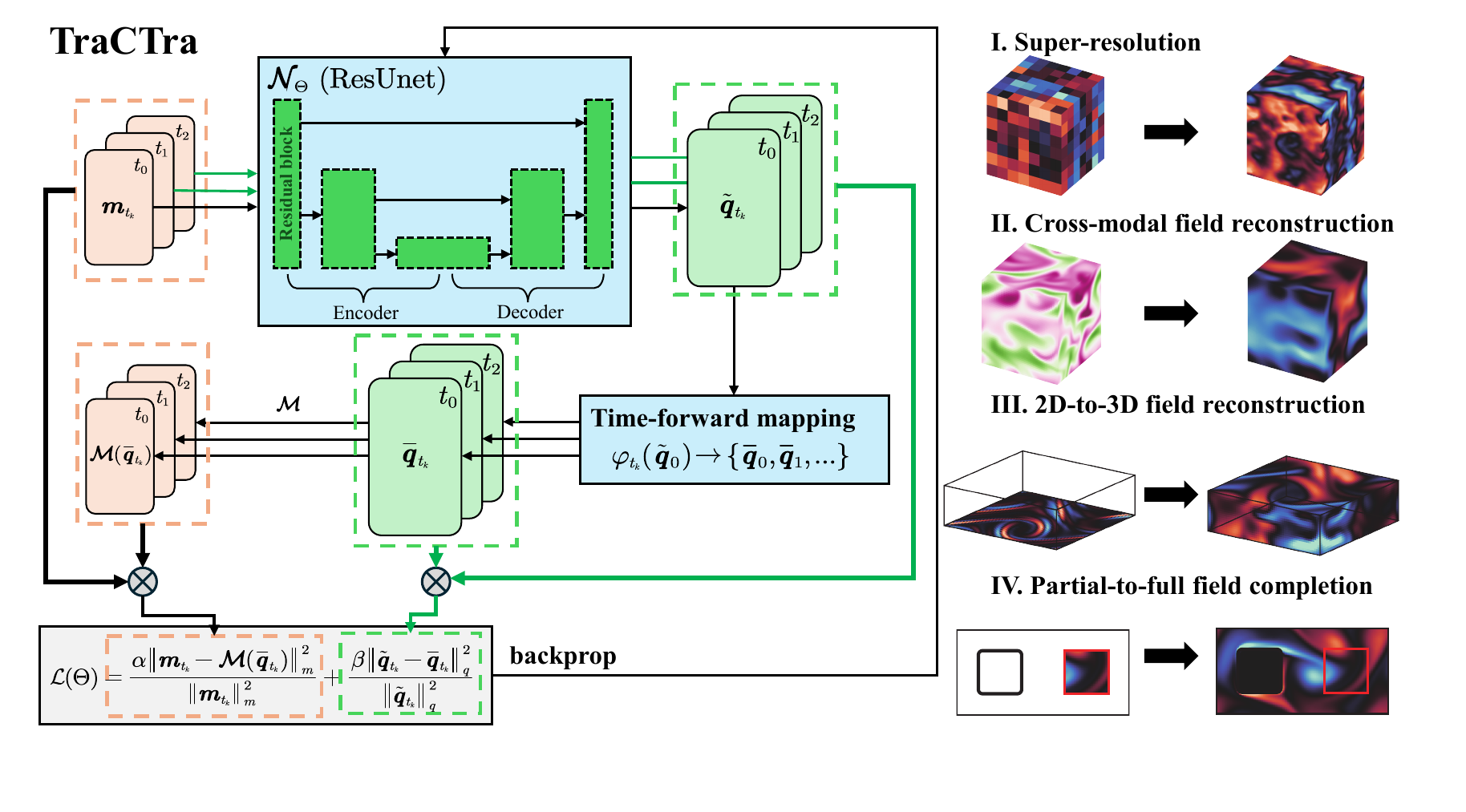}
\caption{\JP{Overview of the algorithm and physical problems studied in this paper}. (Left) \JP{Schematic of the training loop which operates without the existence of full-state reference data.} Note that backpropagation is performed through both terms in the loss (orange and green boxes). In the diagram, $\bar{\bm{q}}_{t_k}= \boldsymbol{\varphi}_{t_k}(\tilde{\bm{q}}_0)$ denotes the state obtained from the time-forward mapping of the initial condition $\tilde{\bm{q}}_0$ using a differentiable solver. \JP{(Right) T}he four flow-reconstruction tasks considered in this paper.}\label{fig:workflow}
\end{figure}

The primary aim of \JP{TraCTra} is to reconstruct a physical flow field from sparse, incomplete, or masked observations. The approach is inspired by 4DVar~\citep{foures2014data,wang2021state}, in which an initial condition is optimized so that its time-evolved trajectory matches the available observations.
\JP{In this spirit, our algorithm assumes that data is available in the form of measurement time series, each of the form $\{\mathbf m_{t_k} \}_{k=0}^{N_T}$, where the measurements will generally consist of far fewer elements than the dimension of the state vector, $\mathbf q$. 
We consider a variety of possible measurements here, motivated by what is available in an experiment, including spatially-coarse observations of velocity, observations of the flow in local patches of the full domain and observations that cannot be directly related to the full velocity field in any simple manner (e.g. density shadowgraph measurements).
}

\JP{
The objective in each problem is to train a neural network which returns an estimate of the full, instantaneous state at time $t$ given a measurement snapshot or short sequence of snapshots, $\tilde{\mathbf q}_{t} := \mathcal N_{\Theta} (\mathbf m_t)$ or $\tilde{\mathbf q}_{t_k} := \mathcal N_{\Theta} (\{\mathbf m_{t_{k+i}}\}_{i=-K}^{K})$ (here the neural network is $\mathcal N_{\Theta}$ and the subscript $\Theta$ denotes its parameters; a tilde will always signify an network prediction; a short, concurrent time series of $2K+1$ snapshots is sent to the neural network -- in most cases $K=0$ and a single snapshot is used).
We are concerned with cases in which no high resolution reference data for the state is available, and thus the neural network must learn to estimate $\mathbf q$ by comparing time-evolved measurement predictions only. 
}

\JP{
The most natural way to accomplish this is to simply use the classical 4DVar loss function to train the network \cite{page2025super,weyrauch2026state}:
}
\JP{
\begin{equation}
    \mathcal{L}(\Theta)=
    \sum_{k=1}^{N_T}\frac{\| \mathbf m_{t_k} - \mathcal M(\boldsymbol \varphi_{t_k} \circ \mathcal N_{\Theta}(\mathbf m_{t_0}))\|_m^2}{\|\mathbf m_{t_k} \|_m^2} = 
    \sum_{k=1}^{N_T}\frac{\| \mathbf m_{t_k} - \mathcal M(\boldsymbol \varphi_{t_k}(\tilde{\mathbf q}_0))
    \|_m^2}{\|\mathbf m_{t_k} \|_m^2}.
    \label{eqn:loss_4dvar}
\end{equation}
Here the loss is written for a single trajectory for clarity.
Below we do train models to estimate the full state with observations extracted from one trajectory (a PINN-like inverse problem), but 
typically break the observations up into $N_S$ shorter `subtrajectories' to avoid backward gradient propagation over large times (well beyond the Lyapunov time for the systems considered). 
The loss in this case reads
\begin{equation*}
        \mathcal{L}(\Theta)=
    \sum_{j=1}^{N_S}\sum_{k=1}^{N_T}\frac{\| \mathbf m_{t_k}^j - \mathcal M(\boldsymbol \varphi_{t_k}(\tilde{\mathbf q}^j_0))
    \|_m^2}{\|\mathbf m_{t_k} \|_m^2},
\end{equation*}
with the index $j$ a label for the subtrajectory.
With the loss written in this way a more general model can be also sought by summing over a large number of \emph{independent} short trajectories (see \S\ref{sec:par-full}).
In equation (\ref{eqn:loss_4dvar}) the network makes a prediction of the full state at $t=0$ given only the initial measurement $\mathbf m_{t_0}$. 
The state is then marched forward in time using a differentiable solver, $\boldsymbol \varphi_{t_k}(\tilde{\mathbf q}_0)$. 
An observation operator $\mathcal M$ then converts $\boldsymbol \varphi_{t_k}(\tilde{\mathbf q}_0)$
into measurements for comparison to the measurement data; the subscript $m$ on the norms indicates that these comparisons are made in the measurement or observation space.
In some of the examples, the network will be given a short time sequence of measurements, $\{\mathbf m_{t_{k-K}}, \dots, \mathbf m_{t_k},\dots, \mathbf m_{t_{k+K}}\}$, which are treated as separate channels in an input `image' when input to the neural network.
}

By backpropagating through the solver, \JP{minimization of} the loss function updates the \JP{neural network parameters to enable prediction of the initial condition} \JP{$\tilde{\mathbf q}_0$} such that \JP{measurements extracted from its forward-time} trajectory are consistent with the observations \JP{-- see the black pathway in figure \ref{fig:workflow}}. 
However, such consistency over a finite unroll time does not guarantee that the inferred initial condition is unique, physically meaningful, or close to the true state. When observations are sparse or strongly masked, they may be insufficient to fully constrain the optimization landscape, making a global optimum difficult to identify. Instead, the optimization may converge to a local minimum that is dynamically consistent with the observations over the assimilation window but remains physically unrelated to the true flow state.

\subsection{\JP{Full-state consistency}}
To address this limitation, we introduce an additional training pathway, shown by the green path in figure ~\ref{fig:workflow}. 
\JP{The assimilation approach above uses the neural network only to predict an initial condition, which ignores the rich information contained in the other measurements in the time series. 
We therefore modify the approach so that the network is also required to predict the full state from the later-time measurements, which are compared to the trajectory time-marched from the network-predicted initial condition. 
This self-consistency contribution is identified by the green lines in figure \ref{fig:workflow}, and the associated loss function reads:
}
\JP{
\begin{align}
    \mathcal{L}(\Theta) &=
    \sum_{k=1}^{N_T}\left( \alpha\, \frac{\| \mathbf m_{t_k} - \mathcal M(\boldsymbol \varphi_{t_k} \circ \mathcal N_{\Theta}(\mathbf m_{t_0}))\|_m^2}{\|\mathbf m_{t_k} \|_m^2} 
    + 
    \beta\, w_{t_k} \, \frac{\|\mathcal N_{\Theta}(\mathbf m_{t_k}) - \boldsymbol \varphi_{t_k} \circ \mathcal N_{\Theta}(\mathbf m_{t_0}))\|^2_q}{\|\mathcal N_{\Theta}(\mathbf m_{t_k}) \|^2_q} 
    \right) \nonumber \\
    &=
    \sum_{k=1}^{N_T}\left( 
    \alpha\, \frac{\| \mathbf m_{t_k} - 
    \mathcal M (\boldsymbol \varphi_{t_k}(\tilde{\mathbf q}_0)) 
    \|^2_m}{\|\mathbf m_{t_k} \|^2_m}
    + 
    \beta\, w_{t_k} \, \frac{\|\tilde{\mathbf q}_{t_k} -
    \boldsymbol \varphi_{t_k}(\tilde{\mathbf q}_0)
    \|^2_q}{\|\tilde{\mathbf q}_{t_k}\|^2_q} 
    \right).
    \label{eqn:loss_preda}
\end{align}
The new contribution, preceded here by the hyperparameter $\beta$, makes comparisons in the full state space (note the subscript $q$ on the norms) but these high-fidelity predictions are made by the network: no aspect of the method relies on the existence of high-resolution reference data. 
The time-dependent weights in the new term, $w_{t_k} = \exp(t_k / T_e)$ ($T_e$ is the unroll window length) place greater emphasis on later times in the assimilation window.
The results in this paper were all obtained with $\alpha = \beta = 1$. 
Again, we have written the loss for a single trajectory for clarity, but in practice break the observations into $N_S$ shorter subtrajectories.
}

In this study, we primarily employ a U-Net with residual convolutional blocks~\citep{zhang2018road} \JP{to construct the full state from measurement snapshots} (see section \ref{sec:resunet}). 
\JP{However, unlike} \JP{many machine learning approaches which are tied to specific architectures} (\JP{e.g. PINNs which are typically} implemented using \JP{fully connected networks}), TraCTra is not tied to any specific \JP{neural network architecture.}
It is also worth noting that the observation operator $\mathcal{M}$ in fig.~\ref{fig:workflow} is a general mapping and is not restricted to a particular form. In the present study, $\mathcal{M}$ represents different operations depending on the reconstruction task: a pooling operator, which maps the high-resolution truth to coarse observations; a masking operator, which removes all/part of the velocity field or the vorticity field; and a transformation operator, which maps a 3D density field to a 2D shadowgraph representation.

\section{\JP{State estimation}}\label{sec:state_est}

\JP{In this section we apply the TraCTra algorithm to four state-estimation tasks. 
All of the examples are derived from fluid mechanics and so the governing equations are either the two- or three-dimensional Navier-Stokes equations or, in cases with density stratification, the Boussinesq equations. 
The specific equations for each case considered below, and relevant changes to the network architecture, are described in Methods.
All equations are solved using a differentiable, spectral Navier-Stokes solver built on the data structures of JAX-CFD~\citep{kochkov2021machine,dresdner2022learning}. Further details are also provided in Methods. 
}

\JP{
In the first two cases, we consider single-trajectory inverse problems, in which the objective is to augment an incomplete measurement vector to estimate the full flow state in a particular short observation window. 
This choice is motivated by the limited data availability typical of practical experiments, as well as by the need to compare the proposed method with other state-estimation approaches that work in this way (4DVar and PINNs).
We then consider a more challenging problem in which measurements are only available on a two-dimensional plane, and consist of a derived quantity (density shadowgraphs); in this case we explore the ability of our model to extrapolate beyond the observed time series.
Finally, we consider a two-dimensional problem where we expand our training to incorporate a set of independent trajectories, and show how TraCTra can be used to train a model which can generalise to unseen data. In both of these last two examples we have been unable to converge classical variational approaches to state estimation, or PINNs. 
}

\subsection{Single-trajectory inverse problems}
\subsubsection{Super-resolution}\label{sec:SR}

\begin{figure}
\centering
\includegraphics[width=0.99\textwidth]{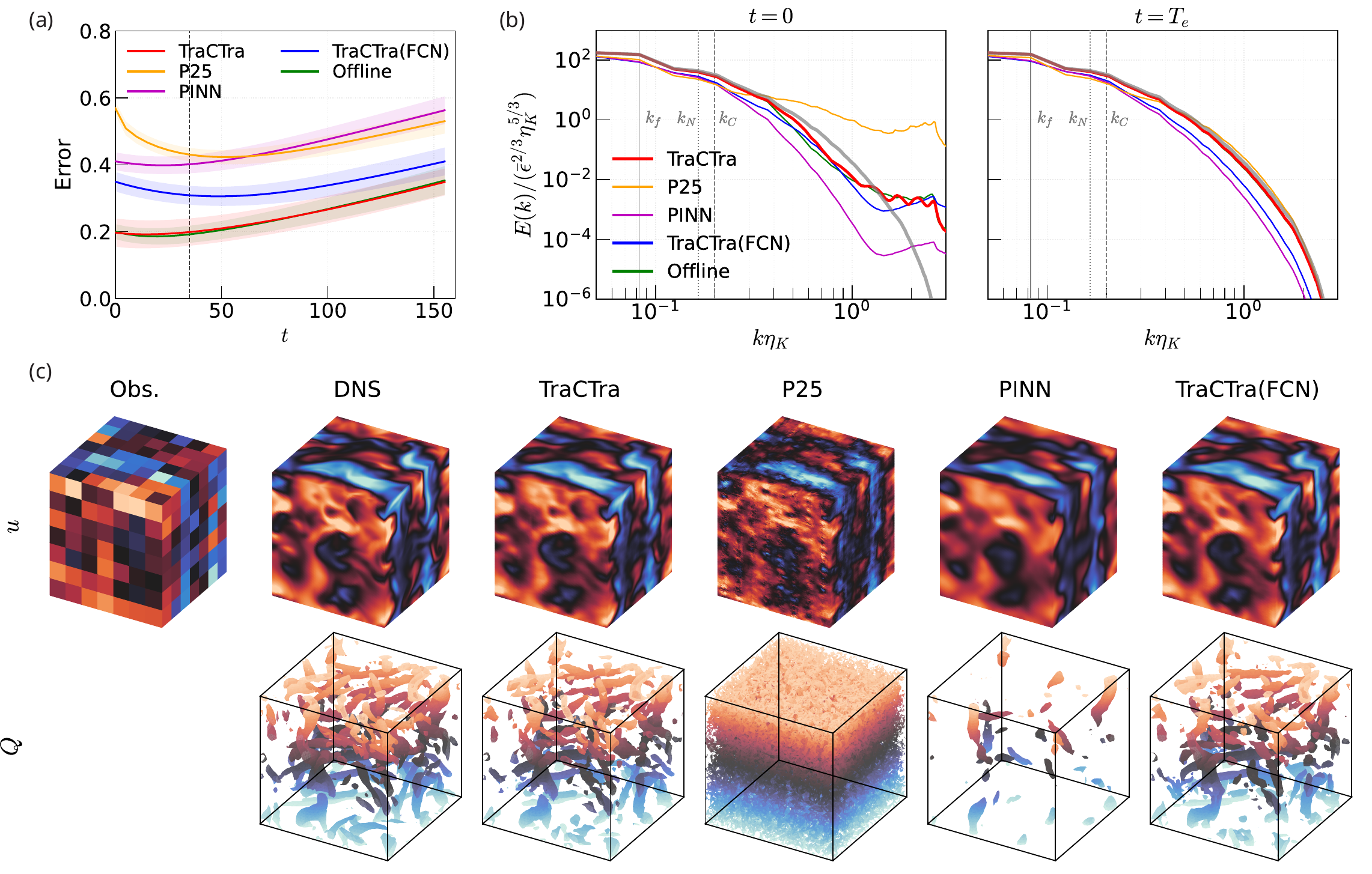}
\caption{Super-resolution \JP{in turbulent, three-dimensional Kolmogorov flow.} 
(a) Relative error, $\JP{\varepsilon(t) := \|\varphi_{t} (\mathbf q_0) - \boldsymbol \varphi_{t} (\tilde{\mathbf q}_{0}))\| /\|\varphi_{t} (\mathbf q_0) \|}$, as a function of roll-out time, with the vertical dashed line marking the unroll time used in training $T_e$, \JP{where $\tilde{\mathbf q}_0$ is an estimated high-resolution velocity field produced by a neural network (see key in figure -- note TraCTra results in bold red)}. Lines are an average over the 200-snapshot dataset, and shaded regions indicate $\pm$ one standard deviation.
(b) Kinetic energy spectra \JP{of the predicted initial condition and after an unroll time $T_e$}, where $k_f$, $k_N$, and $k_C$ denote the forcing wavenumber, the Nyquist wavenumber of the coarse observations, and the synchronization wavenumber, respectively. 
(c) Streamwise velocity $u$ \JP{shown on faces of the domain (top)} and \JP{isosurfaces of constant $Q = Q_\mathrm{rms}$} \JP{(see equation \ref{eqn:q_defn}) coloured by local $z$-value.}}\label{fig:SR}
\end{figure}

\JP{We first consider an inverse problem in which the objective is `super-resolution' of a turbulent flow: prediction of a high-resolution flow field given only spatially coarse velocity data \citep{fukami2019super}.}
\JP{Our turbulence is driven by a monochromatic body force in the $x$ direction, $\mathbf f=Re^{-1} \cos (k_f y)\hat{\mathbf x}$, and we generate data at $Re = 10^4$ \citep[note that the factor of $Re^{-1}$ makes the problem equivalent to unit-amplitude forcing at $Re=100$, see ref][]{Cleary2025}.}
A trajectory covering $1000$ dimensionless time units (TU) is used for training, consisting of $200$ snapshots separated by a time interval $\Delta t=5$. By allowing overlap between training windows, approximately $200$ subtrajectories are generated and used to train the ResUnet model (fig.~\ref{fig:network}) adopted in the \JP{TraCTra} framework.
\JP{For the purpose of comparison, we also train a model using a supervised offline approach using full-resolution fields, with a mean-squared error objective function}. 
We have also implemented \JP{an alternative, fully-connected} architecture within both PINN and \JP{TraCTra} formulations, where the corresponding spatiotemporal coordinates are used as inputs to predict the associated flow variables. Details of the numerical setup are in the supplementary material.

\JP{We consider an aggressive downsampling of $M=16$, which is coarser than known critical continuous-assimilation lengthscales --} the coarse-grid Nyquist wavenumber $k_N$ lies below the critical synchronization wavenumber $k_C=0.2\eta_K^{-1}$~\citep{Yoshida_Yamaguchi_Kaneda_2005,Lalescu_Meneveau_Eyink_2013}, \JP{where $\eta_K$ is the Kolmogorov length scale}.

\JP{
The performance of a neural network trained using the TraCTra algorithm is examined in figure \ref{fig:SR} alongside various other machine-learning-based state estimators. 
For each network, an initial condition is constructed via $\tilde{\mathbf q}_0 = \mathcal N(\mathbf m_0)$ and marched forward in time; note that at ``test time’’ the model sees \emph{only} the initial measurement vector and makes estimate based on this alone.
In figure \ref{fig:SR}(a) we compare the relative error
\begin{equation}
	\varepsilon(t) := \frac{\|\boldsymbol \varphi_{t} (\mathbf q_0) - \boldsymbol \varphi_{t} (\tilde{\mathbf q}_{0}))\| } {\|\boldsymbol \varphi_{t} (\mathbf q_0) \|},
\label{eqn:rel_error}
\end{equation}
In the time-evolution of the reconstructed fields and the ground-truth trajectory. 
The TraCTra approach (with the ResUNet architecture) leads to robust reconstruction at $t=0$ with $\varepsilon \sim 0.2$, with relative error degrading slightly under time stepping. 
The performance is almost indistinguishable from the offline training approach which was trained with high-resolution reference data, while the performance is $2-3\times$ better than the ``assimilation only’’ algorithm (\ref{eqn:loss_4dvar}) — TraCTra’s self-consistency term (see equation \ref{eqn:loss_preda}) is key. 
}

\JP{
The new algorithm is also a substantial ($\sim 2\times$) improvement over the popular PINN approach (see purple curves in figure \ref{fig:SR}). 
The PINN is constrained by a sub-optimal fully-connected architecture though — although the TraCTra algorithm applied to this less capable predictor still results in an improved performance (blue curves), particular under time marching. 
}

In \JP{the kinetic energy spectra shown in figure \ref{fig:SR}(b)}, \JP{the TraCTra algorithm (see red curve) is an accurate estimator for the energy content in the larger scales above the observation wavenumber, with agreement getting progressively worse at smaller scales, with only the offline, full-resolution baseline model performing comparably.}
\JP{In contrast, all other approaches underestimate the large-scale energy content.}
\JP{At small scales the majority of models} underestimate the small-scale energy content at $t=9$, \JP{while the assimilation only approach (\ref{eqn:loss_4dvar})} distinctly overestimates it. 
At $k>k_C$, \JP{TraCTra} even shows closer agreement with the DNS than the supervised offline baseline \JP{(compare red and green curves)}.

\JP{Unrolling in time (see second panel of spectra in figure \ref{fig:SR})(b)) shows that the errors in the tails of the TraCTra (ResUNet) and the offline spectra rapidly collapse to the DNS ground truth, while the other approaches (particularly those involving fully connected networks, including the PINN) retain noticeable differences even in the largest scales. }

\JP{The instantaneous snapshots of figure \ref{fig:SR}(c)} paint a consistent picture,
\JP{with the TraCTra-trained networks producing outputs which closely resemble the reference DNS.}
\JP{The self-consistency term \eqref{eqn:loss_preda} yields smooth fields which yield realistic reconstructions of derived quantities: here we show isosurfaces of the second invariant of the velocity gradient \citep[the $Q$ criterion][]{hunt1988eddies},}
\begin{equation}
    Q=\frac{1}{2}\left(\|\boldsymbol{\Omega}\|^2-\|\mathbf{S}\|^2\right),
    \label{eqn:q_defn}
\end{equation}
where $\boldsymbol{\Omega}$ and $\mathbf{S}$ are the antisymmetric and symmetric parts of the velocity-gradient tensor, respectively.
\JP{The success of TraCTra can be contrasted to the high-wavenumber noise-dominated fields from the assimilation-only approach and the over-smoothed output of the PINN.}

\subsubsection{\JP{Estimating velocity fields from density measurements}}\label{sec:CM}

\begin{figure}
\centering
\includegraphics[width=0.98\textwidth]{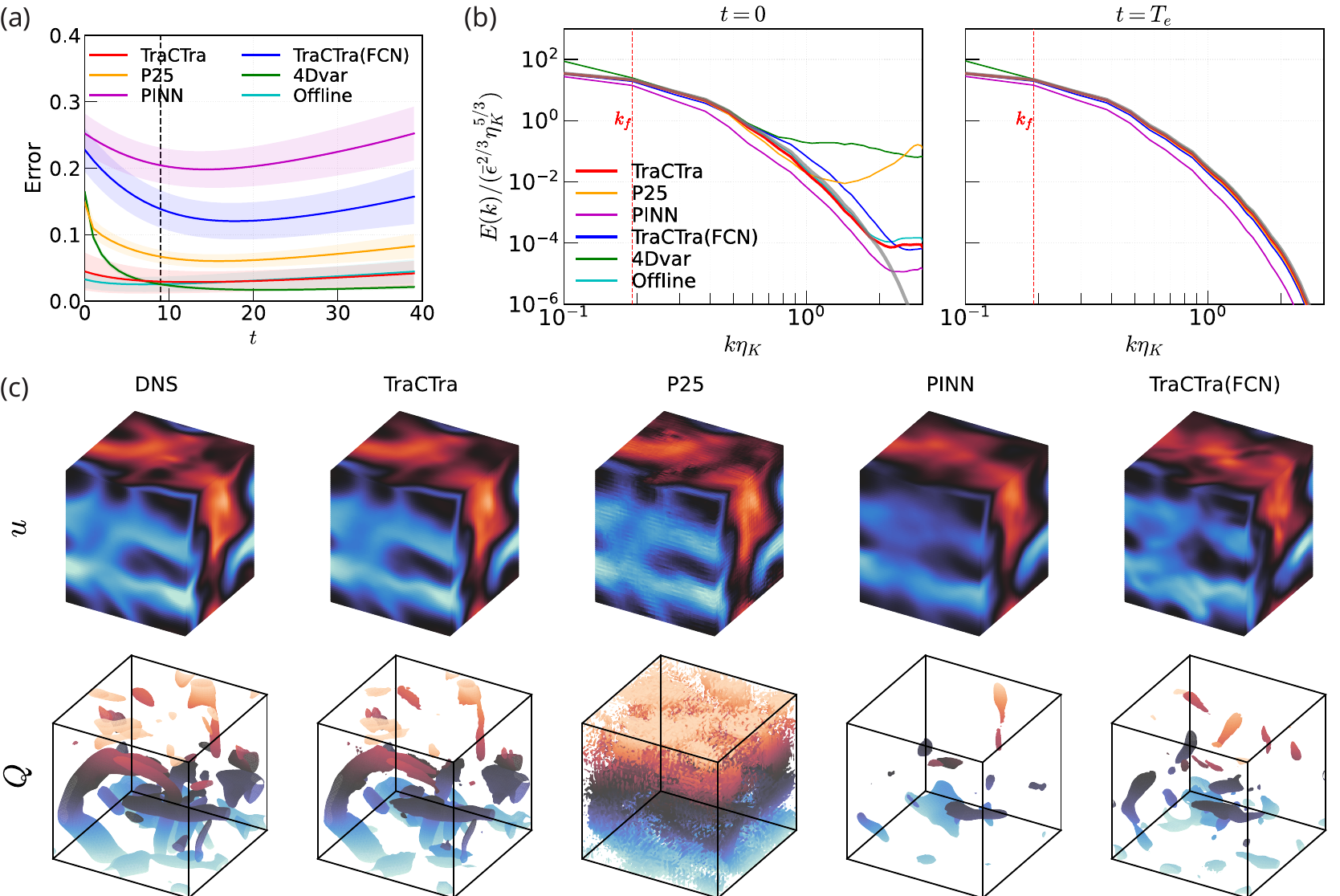}
\caption{\JP{Velocity estimation from density measurements for various estimators.}
(a) Relative error $\varepsilon(t)$ \JP{(equation \ref{eqn:rel_error})} as a function of rollout time. Lines denote averages over all velocity and density components across the 200-snapshot dataset, and shaded regions are $\pm$ one standard deviation.
(b) Kinetic energy spectra. 
(c) \JP{An example snapshot of true (DNS) and estimated} streamwise velocity $u$ and $Q$ vortex fields. 
\JP{Note that the `4DVar' results in panel (a) are the average of five independent computations, each performed with rollout time $T_e=10$.}}
\label{fig:CM}
\end{figure}

Beyond resolution limitations, many fluid-flow applications involve quantities that are difficult to measure directly owing to limited or impractical diagnostic techniques. In stratified flows, for example, simultaneous measurements of velocity and density typically require combined PIV/PLIF systems, involving additional cameras, optical filters, calibration procedures, and careful synchronization~\citep{dossmann2016mixing,partridge2019versatile}. This substantially increases the cost, complexity, and setup time of the experiment. 
\JP{In this section we consider the problem of estimating the velocity field given only density measurements, where the TraCTra training algorithm does not require the existence of reference velocity data. }

\JP{For illustration we consider} 3D triply-periodic Kolmogorov flow \JP{with uniform (stably stratified) background density gradient (see Methods), similar to that described in \citet{cope2020dynamics}.} 
\JP{As in the super-resolution problem above, we drive the flow with a monochromatic force $f_x=Re^{-1}\sin(y)$ in a box} of size $(2\pi)^3$. 
The simulations are performed at $(Re,Pr,B)=(500,1,0.001)$ on a $64^3$ grid\JP{, where $B$ is the buoyancy coefficient in the dimensionless momentum equation (see Methods for discussion of the relation of our parameters to those with a unit-amplitude forcing)}. 
Measurements are provided from a trajectory spanning $200$ time units with $200$ snapshots ($\Delta t=1$), which yields
approximately $190$ overlapping subtrajectories \JP{each of length $T_e=10$}. 

\JP{
In this problem no algorithm is given access to reference velocity data (other than the offline approach included for the purposes of comparison). Note that the PINN and the fully-connected TraCTra approach are trained in a different manner with coordinates $(x,y,z,t)$ as input, see Supplementary Information.
We also include some results from standard variational assimilation (the 4DVar algorithm) which seeks an initial condition via optimization against an individual measurement time series \citep{Wang_Zaki_2021,ZakiARFM2025},
\begin{equation}
	\mathcal L_{\text{4DVar}}(\mathbf q_0) := \sum_{t_k} \| \mathbf m_{t_k} - \boldsymbol \varphi_{t_k} (\mathbf q_0)\|^2.
\end{equation}
The measurements here consist of the full-resolution density field. In contrast to the super-resolution study above we do not coarse-grain the observations in space in any of our state estimators. 
Specific details of the neural network architectures are provided in the Supplementary Information.
}

\JP{
TraCTra is compared to a variety of alternative state estimators in figure \ref{fig:CM} (see thick red lines).
Similar to the super-resolution problem, its performance is on a par with the reference offline method the absence of any reference velocity data. 
The assimilation-only variant (\ref{eqn:loss_4dvar}) and the fully-connected alternatives are noticeably worse across the comparison window. 
Most striking is that the TraCTra algorithm is competitive with the 4DVar computations at later times; 4DVar also suffers from large early-time errors associated with large-amplitude high-wavenumber noise (see spectra in figure \ref{fig:CM}(b)). 
}

\JP{
For both spectral energy content and coherent structures (figure \ref{fig:CM} panels (b) and (c)), TraCTra is again found to be most consistent with the DNS ground-truth, including the reconstruction of complex $Q$-structures over a range of scales.
The results highlight advantages of the approach over PINN, which retains an under-prediction of energy in all scales over time (see $t=T_e$ energy spectra). 
This is partially an architectural effect (TraCTra with a fully connected network is also poor — though still notably better than the PINN) and highlights the benefits of our training algorithm which is architecture-agnostic. 
}

\subsection{\JP{Within-trajectory extrapolation: full-field reconstruction from 2D density shadowgraphs}}\label{sec:2d-3d}

\begin{figure}
\centering
\includegraphics[width=0.98\textwidth]{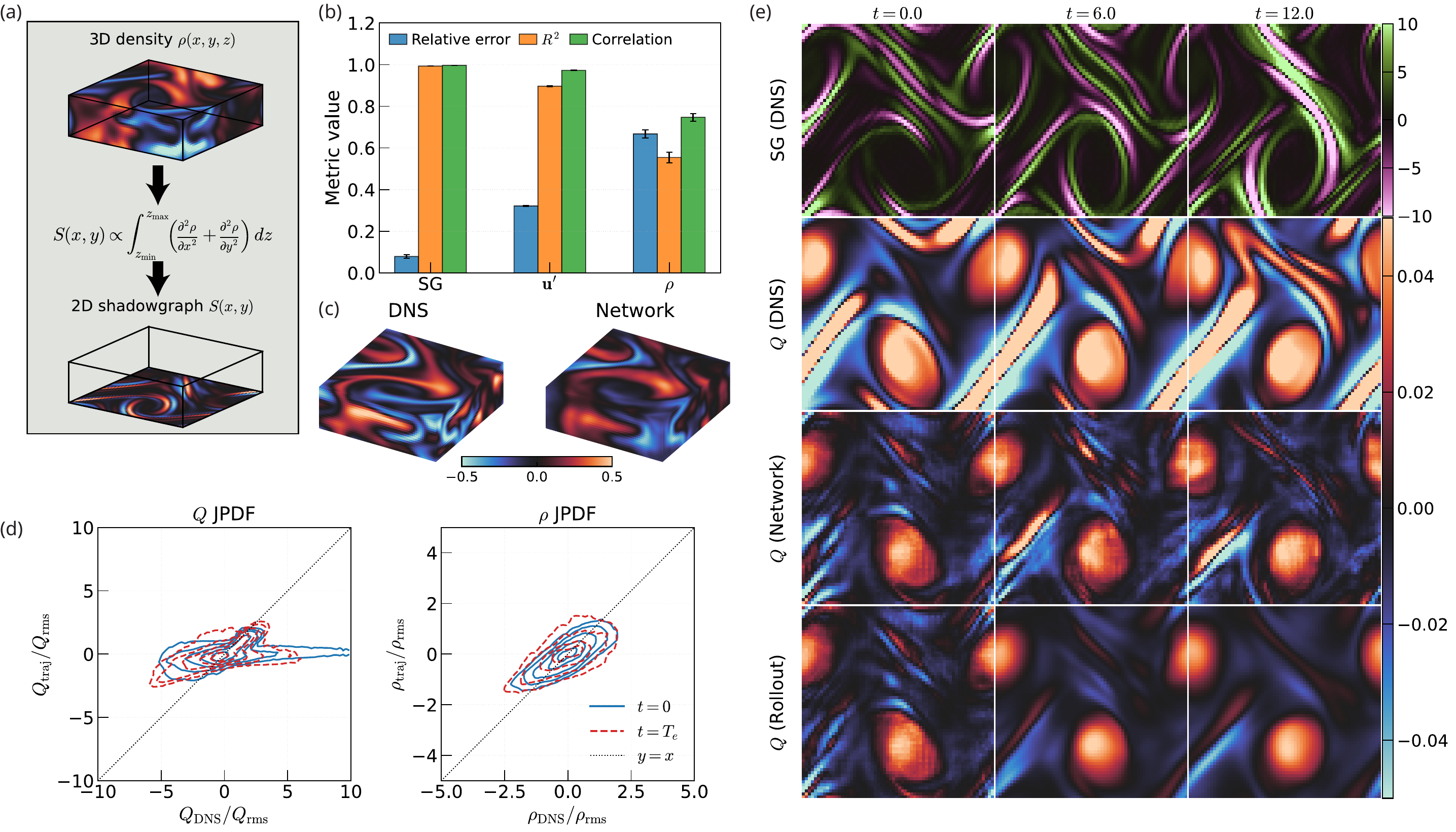}
\caption{\JP{Shadowgraph to full 3D-state reconstruction}. 
(a) Schematic of the shadowgraph computation. \JP{An individual input to the network is a two-dimensional, multi-channel images $\{S[\rho_{t_{k-2}}], S[\rho_{t_{k-1}}], \cdots S[\rho_{t_{k+2}}]\}$ -- a five-point stencil sampling the two adjacent frames (in time) either side of the reference $t_k$ -- the measurements from which the network seeks to construct the full 3D state}. 
(b) Relative error, $R^2$, and Pearson correlation for the reconstructed fields on a `test' dataset 64 advective time units in length \JP{(extracted immediately after the 600 time-unit training trajectory)}.
(c) \JP{A comparison of a} three-dimensional density field \JP{generated by the DNS and produced by the network from a five time-point stencil of shadowgraph measurements}. 
(d) Joint PDFs of $Q$ and $\rho$ \JP{[We see in the initial condition that high Q events in the DNS can be erroneously mapped to low Q by the network?]}. 
(e) Temporal evolution of the vortical structures identified by the $Q$-criterion \JP{beneath the input shadowgraph}; \JP{here `rollout' indicates the network was used to estimate the initial field, which was then time marched by the solver, while `network' indicates the snapshot at each time was produced by inputting the corresponding shadowgraph measurement into the solver}. }\label{fig:2d-3d}
\end{figure}

A more challenging reconstruction problem is to recover 3D flow fields from 2D flow projections/slices. 
In experiments, image-based velocimetry techniques such as planar PIV are typically restricted to 2D measurement planes~\citep{adrian1991particle}. Although volumetric techniques such as tomographic PIV can recover 3D velocity fields from multi-camera images~\citep{elsinga2006tomographic}, they require more complex optical setups and reconstruction procedures, with accuracy often constrained by optical access, seeding density, and measurement volume. This motivates data-driven reconstruction of 3D flow fields from simpler 2D observations.
Here, we consider the reconstruction of a 3D stratified flow from 2D shadowgraph images, as illustrated in fig.~\ref{fig:2d-3d}(a). Shadowgraphy (SG) is widely used in experimental fluid dynamics to visualize density variations through changes in the refractive-index field~\citep{adrian1991particle}. 
Under suitable approximations, the shadowgraph signal can be related to the line-of-sight integral of the transverse curvature of the density field,
\begin{equation}
    S(x,y)\varpropto\int_z \left(
    \frac{\partial^2 \rho}{\partial x^2}
    +
    \frac{\partial^2 \rho}{\partial y^2}
    \right)\,{\rm d}z .
\end{equation}
\JP{As} the shadowgraph image is obtained by integration along the $z$ direction, it naturally encodes information from the entire 3D domain \JP{which is beneficial for} the 3D reconstruction. 
The computational domain is $(L_x,L_y,L_z)=(2\pi,2\pi,2)$, discretised using $(N_x,N_y,N_z)=(64,64,20)$ grid points. The governing parameters are $(Re,Pr,\JP{B})=(500,1,0.001)$, \JP{with the physical setup and non-dimensionalisation matching those in the previous section}. 
A trajectory spanning $600$ time units and comprising $300$ snapshots is used for training the network. 
In this problem we supply the network with a short time series of shadowgraph measurements (a stencil of size five centered on the target time $t_k$) which are passed in as a single, five-channel image, from which the objective is construction of the full flow state $\mathbf q_{t_k}$. 
\JP{In contrast to the results above we evaluate our model on a held-out trajectory, 64 time-units in length, extracted after the `training' data (there is no temporal overlap between the datasets). }

Owing to the difficulty of this problem \JP{(the network receives a multichannel image of the derived quantity $S$, without any reference velocity or density fields)}, we report only the \JP{TraCTRa} results \JP{(with ResUNet architecture, see Methods)}, as the other methods generally fail to converge to reasonable solutions \JP{(for reference we include the results of a PINN for the `training-trajectory inverse problem' in the SI)}. 
\JP{TraCTra} uses the shadowgraph images as input to reconstruct the corresponding 3D velocity and density fields. 
The shadowgraph signal computed from the rolled-out trajectory is then compared with the reference shadowgraph.

In figure \ref{fig:2d-3d}(b), we assess the reconstruction performance using the relative error, $R^2$, and Pearson correlation between the model prediction and DNS for the SG signal, velocity fluctuation $\bm{u}^\prime$, and density $\rho$. 
\JP{Reconstruction of both the shadowgraph signal and the unknown 3D velocity field is relatively robust, though the density field is more challenging.}
This is \JP{perhaps to be} expected because the shadowgraph signal measures the line-of-sight integral of the transverse density curvature and therefore does not uniquely constrain the magnitude or depth-wise distribution of the 3D flow field. 
Nevertheless, the relatively high correlations for both $\mathbf u^\prime$ and $\rho$ indicate that the main flow structures are captured. 
This is confirmed in figure \ref{fig:2d-3d}(c), where the dominant density structures are recovered, although their amplitudes differ from the DNS.

The joint probability density functions \JP{reported in figure \ref{fig:2d-3d}(d)} provide a more detailed statistical comparison: 
For both $Q$ and $\rho$, the distribution is tilted counterclockwise away from the one-to-one line, indicating systematic underestimated amplitude errors in the reconstructed vortical and density fields. Temporal rollout to $T_e = 16$ improves the reconstruction of $Q$ but is less pronounced on $\rho$. 
\JP{We also} compare the temporal evolution of the vortical structures in figure \ref{fig:2d-3d}(e).
\JP{Here we show two-dimensional slices of $Q$} obtained from the DNS, \JP{along with} the prediction \JP{of $Q$ computed from a multichannel (five-point stencil in time, see above) shadowgraph measurement, $\mathcal N(\{S\})$}, and the \JP{$Q$ field obtained after} solver rollout \JP{of a network-predicted initial condition}. 
The network and rolled-out trajectories reproduce the main structural evolution, but their amplitudes remain slightly underestimated. The network prediction also contains small-scale artifacts and reduced smoothness. These artifacts are rapidly reduced after a short solver rollout.

\subsection{Generalisation to new trajectories: partial-to-full field reconstruction}\label{sec:par-full}

\begin{figure}
\centering
\includegraphics[width=0.99\textwidth]{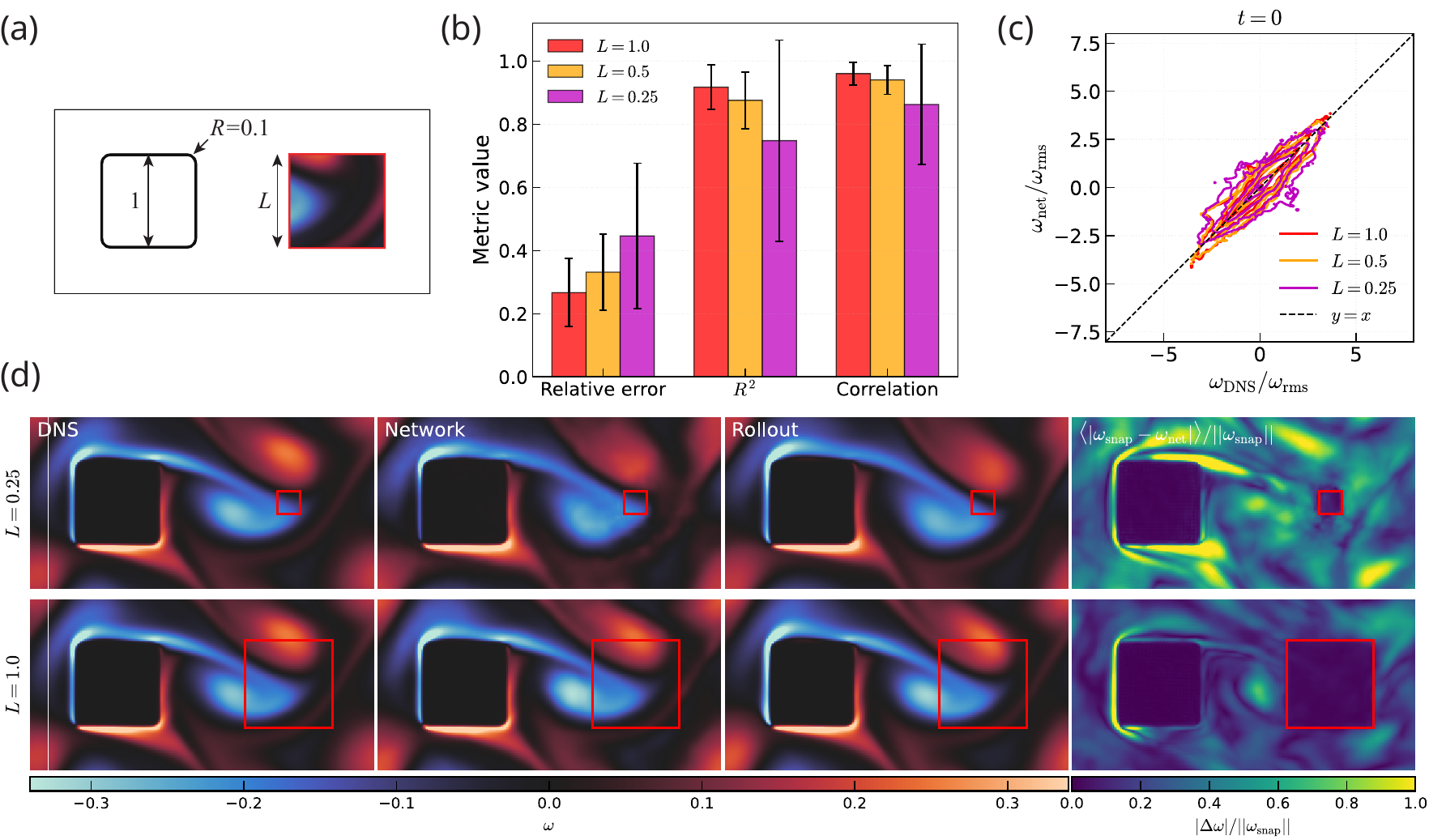}
\caption{Partial-to-full field \JP{reconstruction}.
(a) \JP{Schematic of} flow geometry \JP{and observation window}. 
(b) Relative error, $R^2$, and Pearson correlation for the reconstructed fields \JP{for varying observation window size $L\in\{0.25, 0.5, 1\}$}. 
(c) Joint PDF of \JP{true vorticity against network predictions of the same quantity based on instantaneous measurements}.
(d) \JP{Reconstruction of an example out-of-plane vorticity field for $L\in\{0.25,  1\}$, comparing DNS ground-truth at a time $t=100$, an instantaneous network prediction based on observations in the red window and a time-marched state $\boldsymbol \varphi_T (\mathcal N(\mathbf m_0))$ from a network-estimated initial condition. Final snapshot shows the relative error between the predictions and the DNS ground truth}.}\label{fig:par-full}
\end{figure}

Finally, we examine the capability of \JP{TraCTra} for \JP{full-domain} flow reconstruction \JP{where the measurements consist of observations in a small subdomain}.  
Image completion, or inpainting, is also an important task in flow reconstruction~\citep{buzzicotti2021reconstruction}. It addresses situations where parts of the flow field are unavailable because of limited optical access, finite measurement resolution, or incomplete diagnostics. 
Specifically, we consider partial-to-full image completion of the vorticity field in a 2D flow past an obstacle, with the geometry shown in figure \ref{fig:par-full}(a)\JP{; see Methods for definition of the geometry and handling of the immersed boundary in JAX-CFD}. 
\JP{Unlike the earlier examples which focused on trajectory-specific inverse problems, we train the model here on a set of independent trajectories and evaluate it on completely unseen data. }

\JP{The measurements in this problem are restricted to a}n unmasked square observation region of width $L$ \JP{which} is placed downstream of the obstacle. 
The masked vorticity field, together with the observation mask and geometry function, is provided to the \JP{ResUNet} architecture to predict the full vorticity field.
The computational domain is $(L_x,L_y)=(4,2)$, discretized on a $(N_x,N_y)=(288,144)$ grid. 
\JP{The flow is driven by horizontal body force $f^*$ in the $x$-direction, which we use to define reference a reference velocity scale, $U^* = f^* l^{*2}/\nu$ ($l^*$ is the height of the obstacle); the Reynolds number $Re = U^*l^*/\nu = 5000$.}

Three observation-window sizes are considered: $L=0.25$, $0.5$, and $1$. 
\JP{Similar to the shadowgraph problem above, we provide a short time series of masked observations to the model as a multi-channel image, with stencils of size 9, 5 and 3 (for $L$ increasing from $0.25$ to $1$).}
\JP{Note that the periodic boundary conditions on all edges of the computational domain mean we are considering flow through an array of obstacles; at this particular box/obstacle ratio and $Re$ the dynamics are quasi-periodic}.

To assess the generalisation capability of the model under increased data availability, we train the network on 20 independent trajectories, each spanning 200 time units and containing 25 snapshots. The trained model is then evaluated on a separate test trajectory that is not included in the training dataset
\JP{This approach is feasible here because of the reduced dimensionality of the full state -- the extra consistency term in \eqref{eqn:loss_preda} requires many additional network calls and on-chip storage per batch, something which we discuss further below. }

Figure~\ref{fig:par-full}(b) shows that all cases achieve reasonable accuracy, although the error increases as the observation-window size $L$ decreases, with the strongest degradation observed for $L=0.25$. Consistently, the joint PDFs in figure ~\ref{fig:par-full}(c) show good alignment with the one-to-one line for $L=1$ and $L=0.5$, whereas the broader distribution for $L=0.25$ indicates larger pointwise errors.
The instantaneous fields in figure \ref{fig:par-full}(d) further show that both the $L=0.25$ and $L=1$ cases recover the main vorticity structures. However, the $L=0.25$ case underpredicts high-magnitude vorticity regions. The residuals remain small within the observation window but increase near the obstacle, where the flow is less directly constrained by the available observations. 
A corresponding \JP{assimilation-only (i.e. without TraCTra's consistency term, see equations \ref{eqn:loss_4dvar} and \ref{eqn:loss_preda})} case was also tested but failed to converge to a physically reasonable (see SI).

\section{\JP{Discussion and conclusions}}\label{sec:conclusion}

\JP{
In this paper we have introduced a training algorithm for neural networks to perform state estimation in high-dimensional, chaotic dynamical systems. 
Our method, ``trajectory consistent network training” or TraCTra, combines variational assimilation — matching measurement time series — with a self consistency term which ensures that the networks predictions over the entire measurement time series remain close to the time-marched network output. 
This is done by the inclusion of a (differentiable) flow solver in the loss to advance network predictions forward in time to compare to available measurements -- a much stronger constraint than a soft penalisation based on the residual of the governing equations, for example. 
The unique advantage of these ideas over classical deep-learning techniques is that they do not require the existence of high-resolution, or full-state reference data, making them ideal for new experimental situations where limited information on e.g. boundary conditions or flow parameters may be available, but which could also be determined as part of the problem solution. 
}

\JP{We have considered four specific inverse problems here in which TraCTra outperformed an array of alternative state-estimation approaches.
Our numerical examples were designed to each highlight specific experimental constraints: under-resolution in space, missing dynamical variables, limited observation of derived quantities only and partial observability of the flow domain. 
While some advantage over softer constrained approaches such as PINNs is expected because of our more stringent (and computationally demanding) requirements in the loss, it is also clear the design of the neural network was key in all problems. 
This was clearly exposed above via our inclusion of TraCTra results using simpler fully-connected architectures. However, it also highlights the architecture-agnostic nature of the TraCTra algorithm, which can work with arbitrary network designs.}

\JP{Reliance on a differentiable solver, while important for network performance, is a computational constraint (though, once trained, state estimation on new data requires just a single operation of the model).
There are additional memory costs introduced by the `state consistency' terms (see equation \ref{eqn:loss_preda}): these require many more neural network evaluations -- scaling linearly with number of timesteps and batch size. 
This restriction meant that the earlier examples (super resolution and density-to-velocity) were examined in terms of a single-trajectory reconstruction task, similar to variational assimilation or PINNs. 
We have shown that the method can lead to more general models in our later examples, where the dimensionality of the state vector was reduced (though the actual reconstruction problem was more challenging). 
We would expect our methods to generalise well in the three-dimensional problems too given broader input data, but this would require some further modifications: one candidate would be to reduce the number of full-state comparisons -- perhaps just at the end of the unrolled trajectory -- which we will explore in future work.
Further efforts could also explore to what extent the approach remains functional when the solver is replaced with a cheaper emulator, or whether the differentiable approach can be used to perform a small amount of transfer learning on a larger, more general purpose physics model \citep{mccabe2025walrus}.
}

\section{Methods}\label{sec:method}

\subsection{Neural network architecture}\label{sec:resunet}

\begin{figure}
\centering
\includegraphics[width=0.99\textwidth]{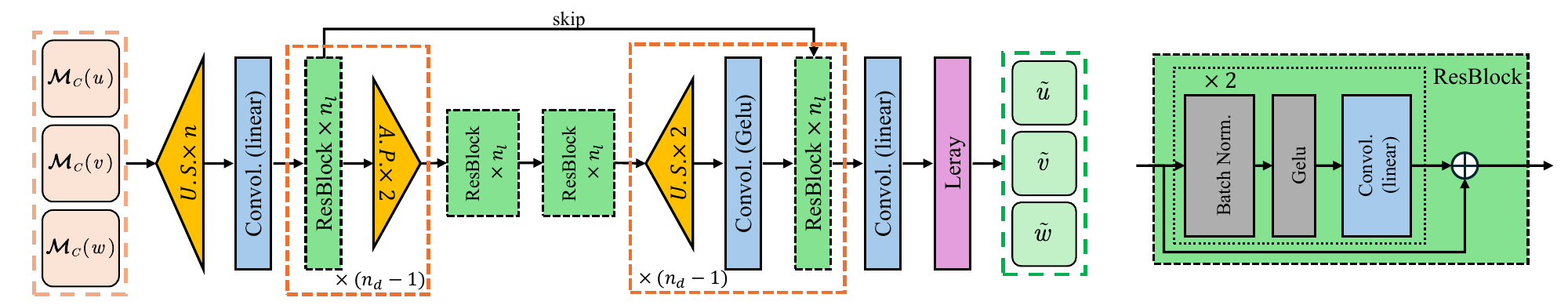}
\caption{\JP{Schematic of the `ResUNet' structure used throughout this work. The ResBlock structure which appears repeatedly in the diagram on the left is shown on the right of the figure. The blocks with `U.S.' and `A.P.' represent nearest-neighbour upsampling and average pooling operations respectively. In the diagram the inputs (for illustration) are coarse velocity measurements, while the outputs are shown as full-resolution velocity fields. The structure of the network and its inputs/outputs differs slightly between problems, as described in the Supplementary Information.}}\label{fig:network}
\end{figure}

The main architecture used in this study is a 3D `ResUnet'~\citep{zhang2018road} \JP{which is sketched} in figure \ref{fig:network} for the super-resolution task \JP{as an example}. 
The coarse input field is first upsampled to the target resolution by a factor $M$ and mapped to $n_f$ feature channels using a periodic convolutional layer.
\JP{The small changes to the architecture required in each individual state estimation task are described in the Supplementary Information.}

The network then follows a U-Net structure~\citep{ronneberger2015u} with residual convolutional blocks. The encoder contains $n_d$ levels \JP{of average pooling (typically $n_d=3\sim5$ depending on the problem, see SI for details of the model's hyper-parameters) interspersed between convolution operations}. 
\JP{Immediately before and after pooling}, the features are processed by $n_l$ residual blocks based on periodic convolutions and the ResNet principle~\citep{he2016deep} \JP{-- see the righter schematic in figure \ref{fig:network} for the structure of an individual residual block}. 
Except at the deepest level, the features are downsampled by average pooling with a factor of two, while the number of channels is doubled. 
The decoder mirrors the encoder. Each decoding level upsamples the feature field by a factor of two, applies a periodic convolutional mixing layer, concatenates the corresponding encoder features through a skip connection, and processes the result with $n_l$ residual blocks. These skip connections help combine coarse contextual information with fine-scale flow structures.

Finally, a periodic convolution maps the decoded features to the velocity components, followed by a divergence-free layer based on the Leray projection,
\[
\mathcal{P}(\bar{\boldsymbol{u}})
=
\bar{\boldsymbol{u}}
-
\nabla \Delta^{-1}\nabla\cdot \bar{\boldsymbol{u}},
\]
which enforces incompressibility of the prediction \JP{exactly without the need for further terms in the loss function}.
\JP{This operation is done in Fourier space where inversion of the Laplacian is a simple division by the square of the total wavenumber.}

\subsection{Problem details}
\label{sec:tf_map}

\JP{In this section we describe the governing equations solved in each of the four examples from the main body of the paper, along with details of the numerical method. We also describe the generation of the training and test datasets.}

In all cases time-forward mapping is carried out using \JP{a suite of DNS codes built on the JAX-CFD software developed by \citet{kochkov2021machine,dresdner2022learning} and modified by \citet{weyrauch2026state} for Fourier-based three dimensional computations.}
For the present applications, we extend these solvers to include a stratified-flow module and a volume-penalization treatment for inhomogeneous flows past immersed objects. The object geometry is represented using signed-distance functions~\citep{osher2003level} and smooth masks.

\subsubsection{Super resolution}
In this problem we work in a triply-periodic computational domain with a monochromatic body force imposed in the streamwise ($x$) direction. 
The velocity is nondimensionalized by the laminar peak velocity $\mathcal{U}$, and spatial coordinates are nondimensionalized by a reference length $\mathcal{L}=L / (2\pi)$, where $L$ is the box height (the box is equal aspect ratio). 
The governing equations are Navier-Stokes and the continuity equation,
\begin{eqnarray}
    \label{eq:ns:mom}%
    \frac{\partial \bm{u}}{\partial t} + \bm{u} \cdot \bm{\nabla} \bm{u} &=&
    - \bm{\nabla}p + \frac{1}{Re} \nabla^{2}\bm{u} + \frac{1}{Re}\cos(k_f\, y)\bm{e}_x,
    \\
    \label{eq:ns:mass}%
   \bm{\nabla} \cdot \bm{u} &=& 0,
\end{eqnarray}
where $Re\equiv \mathcal{U}\mathcal{L}/\nu$ is the Reynolds number and $k_f$ is the forcing wavenumber. 
Note that the use of the laminar velocity to set the velocity scale is the less-common choice -- most authors typically use the forcing amplitude in the definition of a time-scale \citep[e.g.][]{page2025super,weyrauch2026state,cope2020dynamics}. Our Reynolds number can be related to the force-based choice via $Re = Re_{\text{force}}^2$.
\JP{We generate data for the} super-resolution task are at $Re=10^4$ and $k_f=2$ at a resolution of $128^3$.
The numerical solver advances the governing equations in a velocity--vorticity formulation following \citet{kim1987turbulence}, with our differentiable version detailed in \citet{weyrauch2026state}. 
For all problems considered, time integration is performed using a second-order semi-implicit Crank--Nicolson scheme; further details of the numerical implementation can be found in \citet{weyrauch2026state}.

The network inputs are aggressively coarse-grained using a coarse-graining factor of $M=16$, resulting in $8\times 8$ data points per snapshot. The rollout time is set to $T_e=32$, corresponding to approximately $25\%$ of the large-eddy turnover time, as suggested by~\citet{weyrauch2026state}.

\subsubsection{Estimating velocity fields from density measurements}
The cross-modal (velocity from density) flow reconstruction is performed for a three-dimensional stratified Kolmogorov flow. 
The numerical setup and nondimensionalization are similar to those used in the super-resolution case, except that a linear background density stratification is imposed in the $z$ direction, following \citet{cope2020dynamics}. 
The density perturbation is nondimensionalized by the background density variation across the characteristic length scale, $\gamma^* \mathcal{L}$, where $\gamma^*=\partial \rho^*/\partial z^*$ is the dimensional \JP{(constant)} background density gradient and the asterisk denotes dimensional quantities. A Boussinesq approximation is adopted, within which the nondimensional governing equations for the velocity, pressure and density fluctuations about the background state are:
\begin{eqnarray}
    \label{eq:ns:mom}%
    \frac{\partial \bm{u}}{\partial t} + \bm{u} \cdot \bm{\nabla} \bm{u} &=&
    - \bm{\nabla}p + \frac{1}{Re} \nabla^{2}\bm{u} - B\rho\hat{\mathbf z} + \frac{1}{Re}\cos(k_f\, y)\hat{\mathbf x},
    \\
    \label{eq:ns:rho}%
    \frac{\partial \rho}{\partial t} + \bm{u} \cdot \bm{\nabla} \rho &=&
    \frac{1}{Pr\,Re} \nabla^{2}\rho + w,    
    \\
    \label{eq:ns:mass}%
   \bm{\nabla} \cdot \bm{u} &=& 0,
\end{eqnarray}
where $Pr\equiv\nu/\kappa$ is the Prandtl number ($\kappa$ is the thermal diffusivity) and $B \equiv g^*\gamma^*\mathcal{L}^2/(\rho_0 \mathcal{U}^2)$ is a buoyancy parameter (where $g^*$ is the dimensional gravitational acceleration, $\rho_0$ is the reference density). 
The data for this problem are generated at $(Re,Pr,B,k_f)=(500,1,0.001,1)$. 
The DNS setup follows that used for the super-resolution case, with the additional implementation of the background stratification and density-evolution equation. 
The computational domain has size $(2\pi)^3$ and is discretized using a $64^3$ grid.
The rollout time $T_e$ is set to 10.
\JP{Note again the difference in convention in the use of the laminar velocity to define the scaling variables (rather than the forcing amplitude).}

\subsubsection{Full-field reconstruction from shadowgraphs}
The shadowgraph problem is based on the same stratified Kolmogorov-flow system as the density-to-velocity case described above, but with a reduced vertical domain size $L_z=2$ and \JP{correspondingly reduced} $N_z=20$. The rollout time is chosen as $T_e= 16$.

\subsubsection{Partial-to-full field reconstruction}
For the partial-to-full field-completion problem, we consider two-dimensional flow past a periodic array of rounded square obstacles. The computational domain is doubly periodic and represents one repeating unit of the array. Each obstacle has edge length $l^*$ with rounded corners of radius $R^*=l^*/10$ -- the corners are defined as quarter circles  (see figure~\ref{fig:par-full}(a) for the flow geometry). 
\JP{A uniform body force in the horizontal ($x$) direction drives the flow.}
We nondimensionalize lengths by $l^*$ and velocities by the forcing-based viscous velocity scale $U^* \equiv f^* l^{*2}/\nu$ (where $f^*$ is the dimensional forcing magnitude. This scaling gives the following nondimensional governing equations:
\begin{eqnarray}
    \label{eq:ns:mom_IV}%
    \frac{\partial \bm{u}}{\partial t} + \bm{u} \cdot \bm{\nabla} \bm{u} &=&
    - \bm{\nabla}p + \frac{1}{Re} \nabla^{2}\bm{u} + \frac{1}{Re}\bm{e}_x -\frac{\Gamma(\boldsymbol{x})}{\varsigma}\bm{u},
    \\
    \label{eq:ns:mass_IV}%
   \bm{\nabla} \cdot \bm{u} &=& 0,
\end{eqnarray}
where Reynolds number $Re\equiv U^*l^*/\nu$.
The last term on the right-hand of eq.~\ref{eq:ns:mass_IV} is a volume penalty term to impose Dirichlet boundary conditions $\bm{u}=\bm{0}$ that models the submerged object~\citep{schneider2015immersed,hester2021improving}. Here,
\begin{equation}
    \Gamma(x,y) = \frac{1}{2}\left(1+\tanh\left(\frac{2d(x,y)}{\Delta}\right)\right)
\end{equation}
is a smooth mask function for the submerge object, where $d(x,y)$ is a signed distance function~\citep{osher2003level} determining the location of the object. $\Delta$ is a damping length scale, and $\varsigma$ is a damping time scale. Following Hester et al.~\citep{hester2021improving}, we set $\Delta=2\Delta_x$ and $\varsigma=Re(\Delta/2.64822828)^2$ to achieve a second-order numerical accuracy of the volume penalty method.
The simulations are conducted at $Re=5000$ in a doubly periodic domain of size $(L_x,L_y)=(4,2)$, discretized on a $(N_x,N_y)=(288,144)$ grid. 
The governing equations are advanced using the JAX-CFD spectral numerical solver in vorticity formulation~\citep[see][]{kochkov2021machine,dresdner2022learning}. 
\JP{To account for the immersed body we first write the total velocity as $\bm{u}(\mathbf x, t) = U(t)\hat{\mathbf x} + \bm{u}'(\mathbf x,t)$, solving for $\omega = (\boldsymbol \nabla \times \bm{u}')\cdot \hat{\mathbf z}$ and the spatially-averaged mean flow $U(t)$ via}
\begin{eqnarray}
    \label{eq:ns:mom_vor_IV}%
    \frac{\partial \omega}{\partial t} + (\bm{U}+\bm{u}^\prime) \cdot \bm{\nabla} \omega &=&
    \frac{1}{Re} \nabla^{2}\omega -\nabla\times\left[\frac{\Gamma(\boldsymbol{x})}{\varsigma}(\bm{U}+\bm{u}^\prime)\right],
    \\
    \label{eq:ns:U0_IV}%
    \frac{\partial U}{\partial t} &=&\frac{1}{Re}\bigg\langle\frac{\Gamma(\boldsymbol{x})}{\varsigma}(U+u^\prime)\bigg\rangle,
\end{eqnarray}
where $\langle\cdot\rangle$ indicates averaging over both horizontal directions.

\subsection{Training procedure}\label{sec:training}

\JP{The assimilation-based training (see loss \eqref{eqn:loss_4dvar}) is initialised with random network weights, while TraCTra is initialised with a pre-trained assimilation model where one is available.}
This initialization is beneficial because the \JP{neural network `assimilation'} solution already provides a reasonably synchronized solver-evolved trajectory, which can accelerate TraCTra training and reduce the risk of the trajectory-matching term driving the optimization towards an irrelevant dynamical state. 
For more challenging cases, such as the 2D-to-3D and partial-to-full reconstruction problems, the assimilation method does not converge reliably. In these cases, TraCTra networks are trained directly from random initialization.

To improve numerical stability during solver rollout in the loss function \eqref{eqn:loss_preda}, we apply a smooth clipping operation to the network-predicted initial condition before it is passed to the differentiable solver.
If $(\hat{\mathbf q_0})_j$ is the $j^{th}$ component (e.g. the density field) of a neural network output, we transform according to 
\[
(\hat{\mathbf q}_0^{\mathrm{safe}})_j
=
\gamma_n \tanh\left(\frac{(\hat{\mathbf q_0})_j}{\gamma_n}\right),
\]
before time marching.
Here, $\gamma_n$ ($n$ is the training step) is a safety parameter that bounds extreme predictions and prevents unstable solver trajectories during early training. 
The value of $\gamma$ is gradually increased during optimization, typically from $\gamma_0 = O(10^{-2})$ to roughly ten times larger than the maximum value of the corresponding quantity, so that the clipping becomes negligible once the prediction has entered a physically reasonable range.

Finally, in table~\ref{tab:hyperparameters}, we list the hyperparameters used for the optimisation, foward-mapping and data generation of this paper. All models are trained using an Adam optimizer. 

\begin{table*}
    \centering
    \caption{
    Hyperparameters used for the four reconstruction problems:
    $\mathcal{P}_1$, 3D Kolmogorov-flow super-resolution;
    $\mathcal{P}_2$, 3D stratified-flow cross-modal reconstruction;
    $\mathcal{P}_3$, 3D stratified-flow reconstruction from 2D shadowgraph observations;
    and $\mathcal{P}_4$, \JP{partial-to-full field reconstruction.} 
    Here, $l_r$ denotes the learning rate of the optimiser, $B_s$ is the batch size, $T_e$ denotes the forward-mapping unroll time,
    $\Delta t_{\mathrm{obs}}$ is the temporal cadence between consecutive
    snapshots, $N_{\mathrm{sub}}$ is the number of training subtrajectories,
    $T_{\mathrm{burn}}$ is the initial transient discarded during
    training-data generation, $\Delta t_{\mathrm{int}}$ is the time interval between consecutive snapshots along each trajectory. 
    }
    \label{tab:hyperparameters}
    \resizebox{\textwidth}{!}{%
    \begin{tabular}{
        l
        cc
        ccc
        cc
    }
        \toprule
        \multirow{2}{*}{Problem}
        & \multicolumn{2}{c}{Optimizer}
        & \multicolumn{3}{c}{Forward mapping}
        & \multicolumn{2}{c}{Training-data generation}
        \\
        \cmidrule(lr){2-3}
        \cmidrule(lr){4-6}
        \cmidrule(lr){7-8}
        & $l_r$
        & $B_s$
        & $T_e$
        & $\Delta t_{\mathrm{obs}}$
        & $N_{\mathrm{sub}}$
        & $T_{\mathrm{burn}}$
        & $\Delta t_{\mathrm{int}}$
        \\
        \midrule

        $\mathcal{P}_1$
        & $10^{-4}$
        & 1
        & 32
        & 5
        & 193
        & 2000
        & 5
        \\

        $\mathcal{P}_2$
        & $10^{-4}\sim10^{-5}$
        & 5
        & 10
        & 1
        & 191
        & 2000
        & 1
        \\

        $\mathcal{P}_3$
        & $10^{-4}$
        & 10
        & 16
        & 2
        & 289
        & 1500
        & 2
        \\

        $\mathcal{P}_4$
        & $10^{-4}$
        & 10
        & 20
        & 2
        & 385
        & 2000
        & 2
        \\

        \bottomrule
    \end{tabular}%
 }
\end{table*}

\subsection{Code}
All code for the networks, forward solvers and training loops has been made publically available here: \url{https://github.com/luzhu24/TraCTra}

\section*{Acknowledgements}
We gratefully acknowledge support from a UKRI Frontier Guarantee Grant EP/Y004094/1. Computational resources were provided by the Edinburgh International Data Facility (EIDF), the Data-Driven Innovation Programme at the University of Edinburgh.

\section*{Competing interests}
The authors declare no competing interests.

\appendix

\section{Details of neural networks for each state estimation task}\label{sec:SR}

\JP{We describe here specific alterations to the baseline ResUNet architecture outlined in the main text for each state-estimation task. In all cases the network design remains essentially the same aside from changes associated with input/output shapes.}

\subsection{Super-resolution}\label{sec:CM}
The network architecture \JP{for the three-dimensional super resolution task} is shown in fig.~6 of the main text. We use a U-ResNet architecture for the TraCTra algorithm and a fully connected network for PINN. 
The U-ResNet consists of an encoder--decoder structure, each with $n_d=4$ stages of incremental down/upsampling. Each rlevel contains $n_l=2$ residual blocks. The convolutional layers use $(3,3,3)$ kernels, with an initial number of 16 filters that is doubled after each downsampling operation and halved correspondingly during decoding. 
The resulting U-ResNet contains $1,313,267$ trainable parameters.

\subsection{Estimating velocity fields from density measurements}
\begin{figure}[h]
\centering
\includegraphics[width=0.95\textwidth]{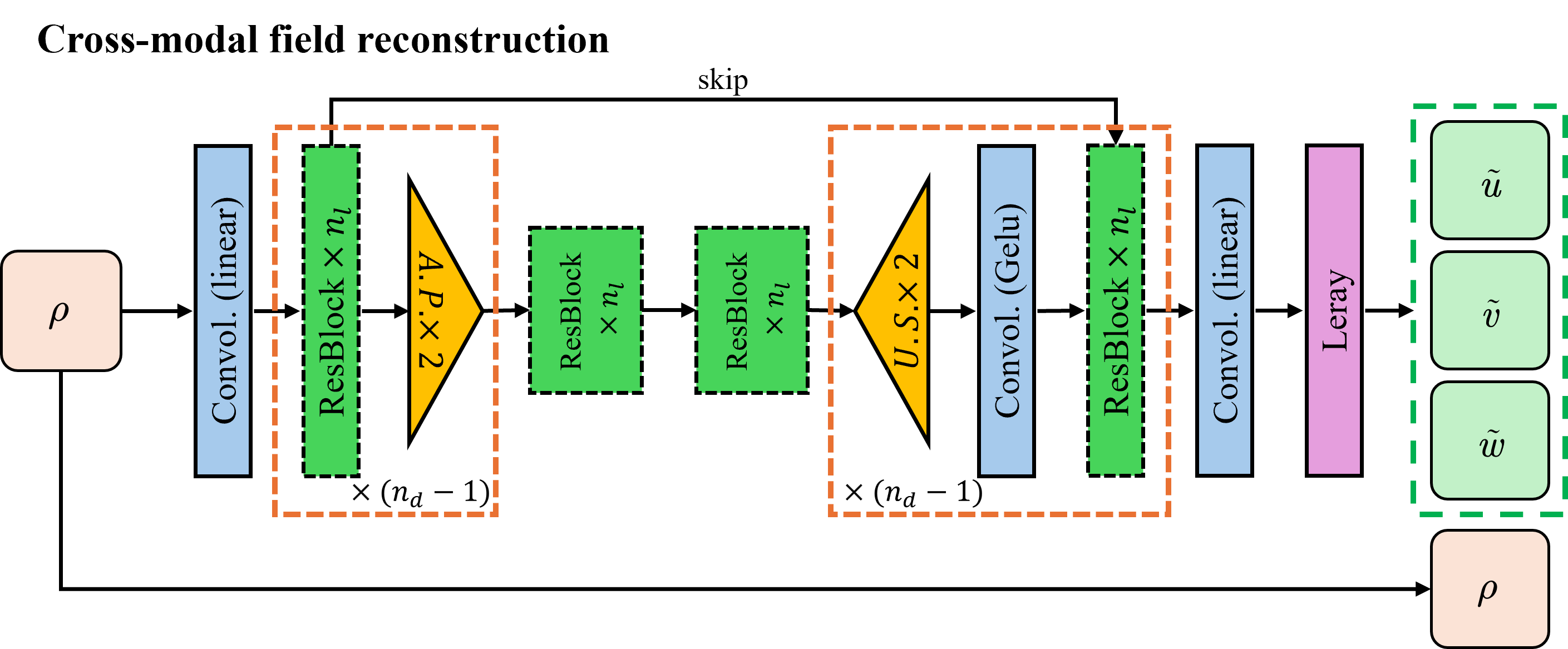}
\caption{U-ResNet architecture for cross-modal field reconstruction. The input is a full-resolution density field.}\label{fig:network_II}
\end{figure}
\JP{The network architectures for TraCTra and the simpler assimilation-only approach for the density-to-velocity problem} are shown in fig.~\ref{fig:network_II}. The density field $\rho$ is passed to the U-ResNet to predict the corresponding velocity fields, which, together with the observed density field, are used to compute the loss functions. The architecture is the same as that used for super-resolution, except that no initial upsampling layer is required. It contains $n_d=3$ resolution levels, with $n_l=2$ residual blocks per level. The initial number of filters is 16, and all convolutional layers use $(3,3,3)$ kernels. The model has $509,395$ trainable parameters.

\subsection{Full-field reconstruction from shadowgraphs}\label{sec:2d-3d}
\begin{figure}[h]
\centering
\includegraphics[width=0.95\textwidth]{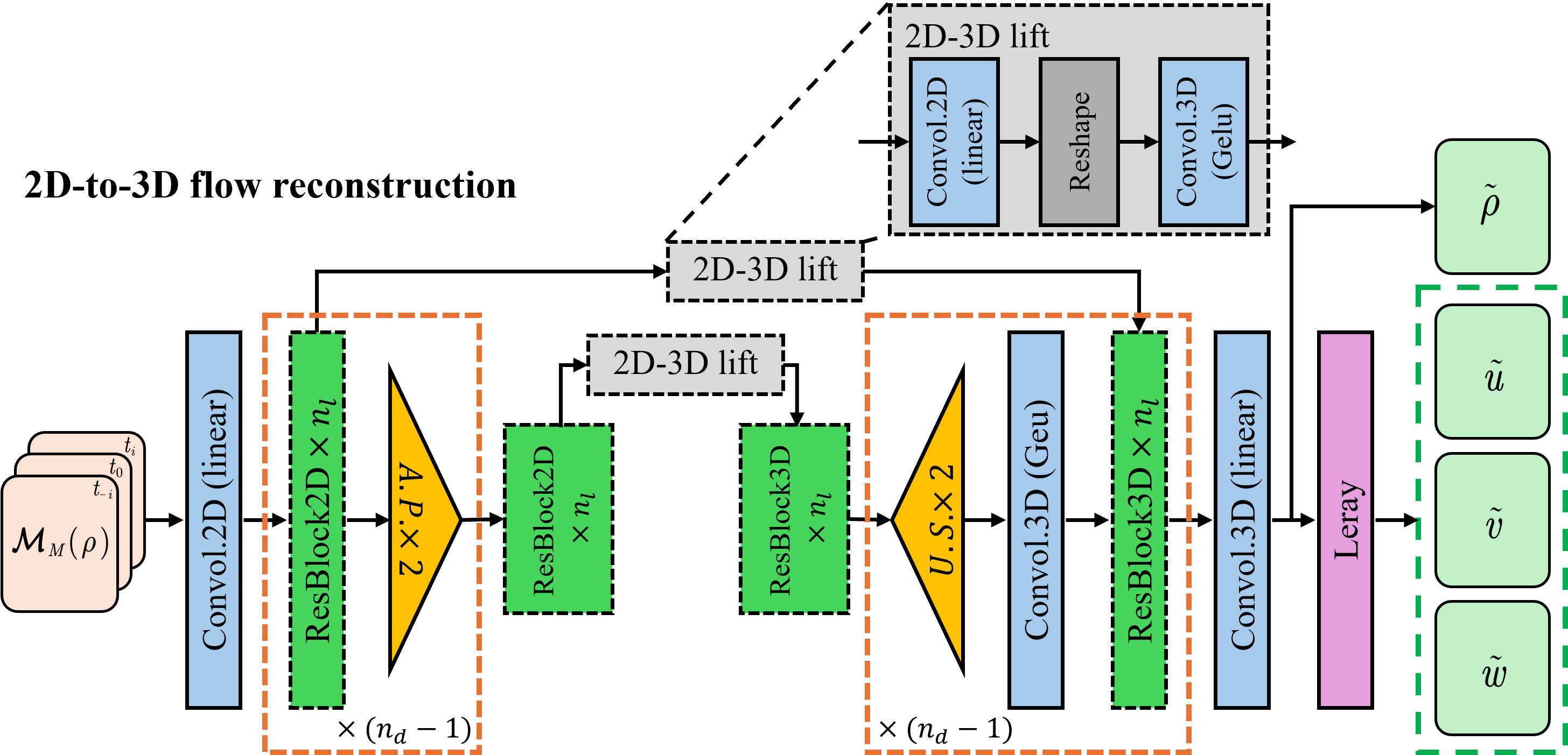}
\caption{U-ResNet architecture for 2D-to-3D reconstruction. The input is a short time series of shadowgraph measurments.}\label{fig:network_III}
\end{figure}
\JP{For the full-state from shadowgraph measurments problem, our} U-ResNet (fig.~\ref{fig:network_III}) uses five successive 2D shadowgraph images, at $t_{i-2},t_{i-1},t_i,t_{i+1}$, and $t_{i+2}$, to reconstruct the 3D velocity and density fields at $t_i$. The input frames are concatenated as channels and processed by a 2D encoder with residual blocks.
At the bottleneck, a learned 2D-to-3D lifting block maps the planar feature representation into a 3D feature volume. This is achieved using a $1\times1$ convolution to generate depth-wise features, followed by reshaping and a lightweight 3D convolution for mixing in the $z$ direction. The skip connections are similarly lifted, allowing multiscale 2D features to be passed to the 3D decoder. The decoder upsamples the lifted features using 3D residual blocks and predicts the velocity and density fields. 
The network has $n_d=3$ resolution levels and $n_l=2$ residual blocks per level. The initial number of filters is 16, with $(3,3)$ kernels in the 2D encoder and $(3,3,3)$ kernels in the 3D decoder. The model contains $726,388$ trainable parameters.

\subsection{Partial to full fields reconstruction}\label{sec:par-full}

\begin{figure}[h]
\centering
\includegraphics[width=0.95\textwidth]{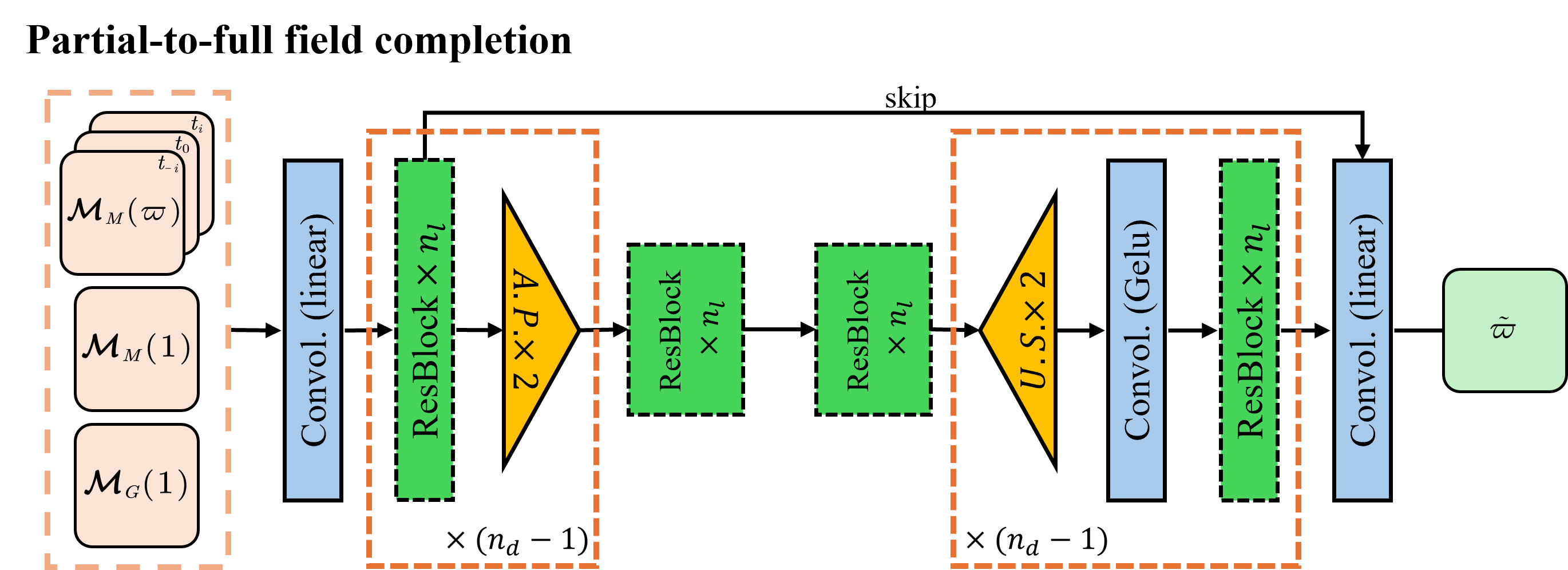}
\caption{U-ResNet architecture for partial-to-full image completion. }\label{fig:network_IV}
\end{figure}

The network architecture \JP{used to estimate a full two-dimensional vorticity field from a small observation window} is shown in fig.~\ref{fig:network_IV}. The input consists of masked vorticity observations at $n_t+1$ consecutive time instances, $\{t_{i}, t_{i+n_t}\}$, with $n_t=4$. The observation mask and immersed-object mask are also included as input channels to provide information about the measurement region and flow geometry.
The network has $n_d=5$ resolution levels and $n_l=1$ residual block per level. The initial number of filters is 16, and the convolutional layers use $(3,3)$ kernels. The model contains $711,281$ trainable parameters.

\section{\JP{Details of fully connected network training}}
\JP{As a comparison to the ResUNet TraCTra approach, we have also included the results from using fully-connected `Physics Informend Neural Networks', or PINNs, as well as a fully connected architecture trained with the TraCTra algorithm. We provide details of these approaches here.}

\subsection{PINN}\label{sec:pinn}

\begin{figure}[h]
\centering
\includegraphics[width=0.95\textwidth]{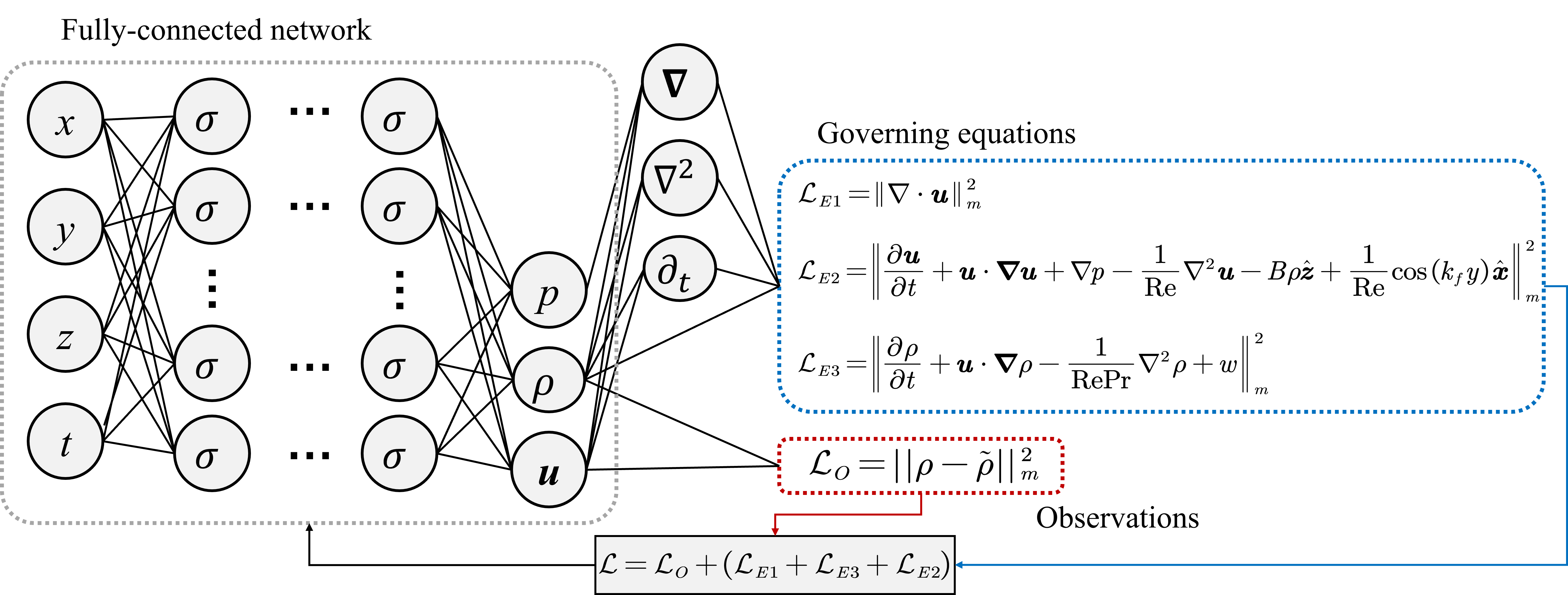}
\caption{Network architecture and procedure of PINN. }\label{fig:network_PINN}
\end{figure}

In both the PINN and TraCTra(FCN) models, we use a fully connected network that takes the spatial and temporal coordinates of individual data points as inputs and returns the corresponding physical quantities, such as $\bm{u}$ and $\rho$, as outputs. A typical network architecture can be found, for example, in \citet{zhu2024new}. Figure~\ref{fig:network_PINN} illustrates the typical PINN procedure, using the three-dimensional stratified Kolmogorov flow, problem 2, as an example.

The outputs of the FCN are compared with the observations and, importantly, are also fed into the corresponding governing equations to compute the equation residuals. These residuals are then used to construct the physics loss, $\mathcal{L}_E$, which regularizes the network. The derivatives appearing in the governing equations are conveniently evaluated using automatic differentiation. For the super-resolution problem, we adopt a 17-layer FCN with neuron numbers
[32, 32, 48, 48, 128, 128, 256, 256, 256, 256, 256, 128, 128, 64, 64, 32, 32],
whereas for the cross-modal reconstruction problem, we use an 11-layer FCN with neuron numbers
[32, 48, 128, 128, 256, 256, 256, 128, 128, 64, 32].
For convenience, the network is evaluated at the coordinates of the fully resolved snapshots, and the physics loss is computed on these coordinate points. The loss function is defined as
\begin{equation}
    \mathcal{L}=\alpha\mathcal{L}_O+\beta\mathcal{L}_E,
\end{equation}

where $\mathcal{L}_O$ and $\mathcal{L}_E$ denote the mean-square errors evaluated over the observation points and equation-sampling points, respectively. An example of the loss definition for the three-dimensional stratified Kolmogorov flow is shown in figure~\ref{fig:network_PINN}. The FCN is trained using the Adam optimizer. The optimization requires approximately 200 and 100 GPU hours on an NVIDIA H100 GPU for the super-resolution and cross-modal reconstruction problems, respectively.

\subsection{TraCTra with FCN}\label{sec:trctra_fcn}

\begin{figure}[h]
\centering
\includegraphics[width=0.8\textwidth]{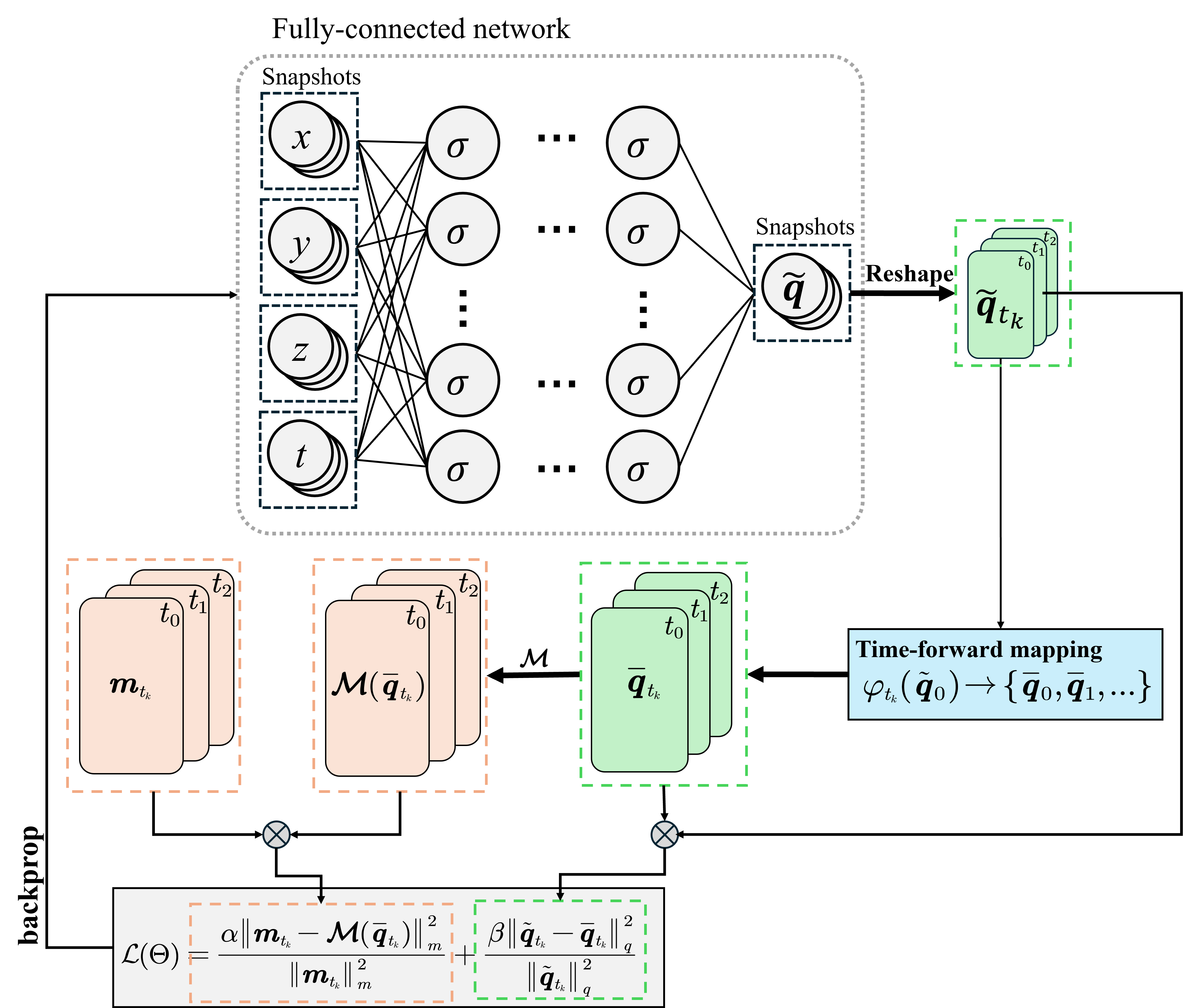}
\caption{Network architecture and \JP{schematic of the training procedure for the TraCTra algorithm with a fully-connected network (`PINN-style')}. }\label{fig:network_pida_fcn}
\end{figure}

The network architecture of TraCTra(FCN) is identical to that of the PINN, while the training procedure is modified to follow the TraCTra framework. At each batch, the coordinates of all snapshots in a subtrajectory are fed into the network to evaluate the state at each point. These pointwise predictions are then reshaped to recover the original trajectory-array structure. The reconstructed state snapshots are subsequently trained following the standard TraCTra procedure. A schematic of the TraCTra(FCN) procedure is shown in figure~\ref{fig:network_pida_fcn}.

For each problem, the network architecture and optimizer are kept consistent with those used for the PINN. The loss function and time-forward mapping are the same as those in the standard TraCTra formulation. The optimization requires approximately 200 and 100 GPU hours on an NVIDIA H100 GPU for the super-resolution and cross-modal reconstruction problems, respectively.

\section{Example failure modes of other estimators}
As described in the main paper, problems 3 and 4 (shadowgraph to full state and masked-to-full field reconstructions) proved challenging for some of the comparison algorithms. 
We include some example outputs here for completeness. 

\begin{figure}
    \centering
    \includegraphics[width=0.5\linewidth]{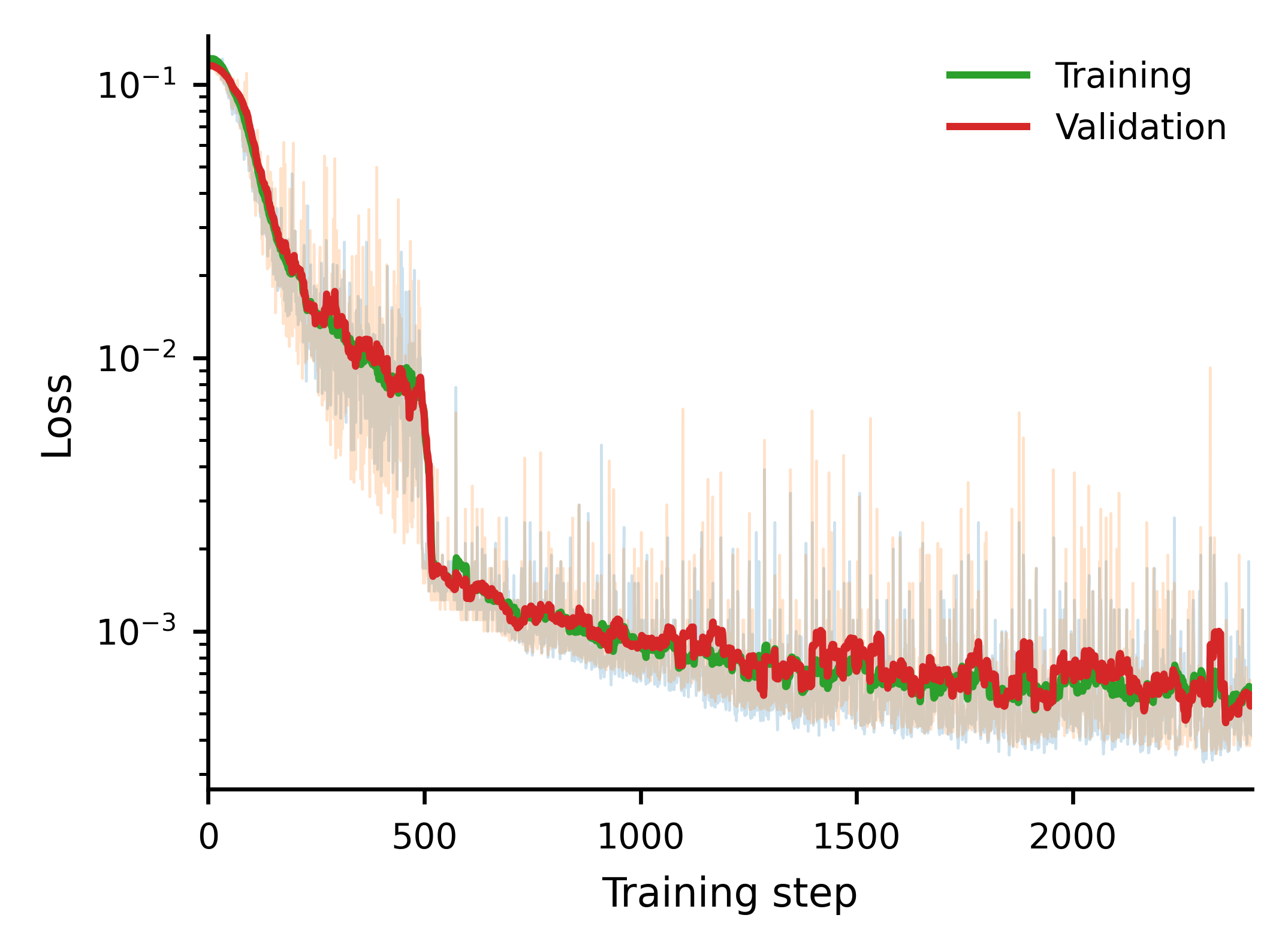}
    \includegraphics[width=0.75\linewidth]{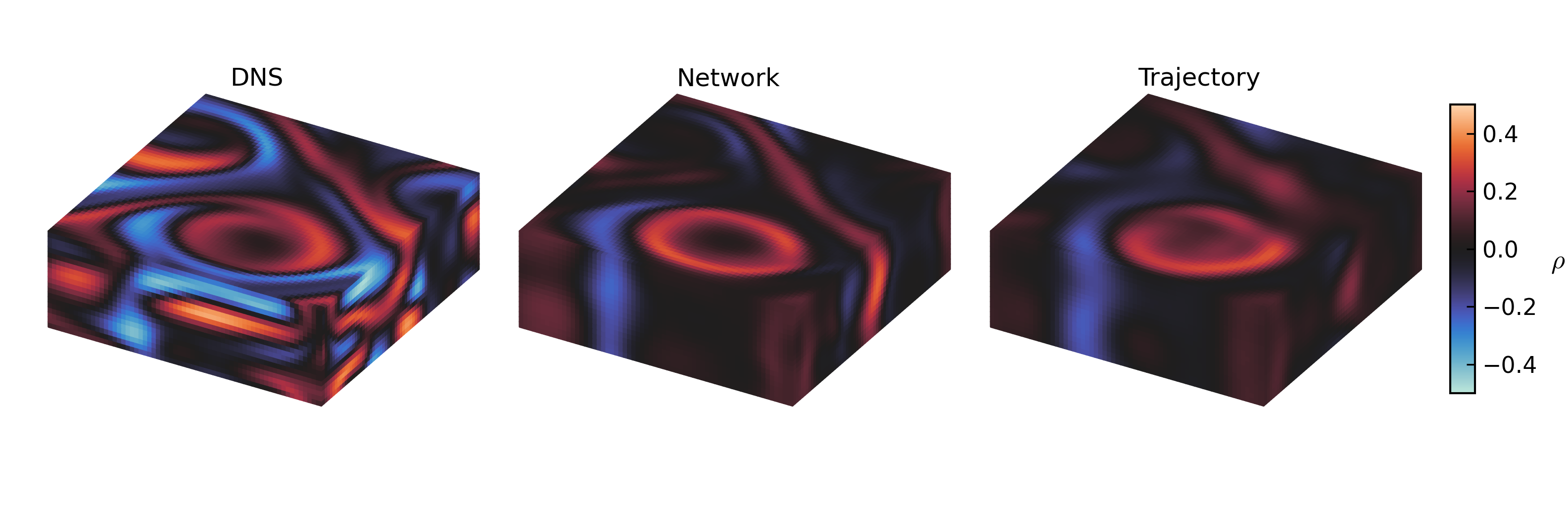}
    \includegraphics[width=0.75\linewidth]{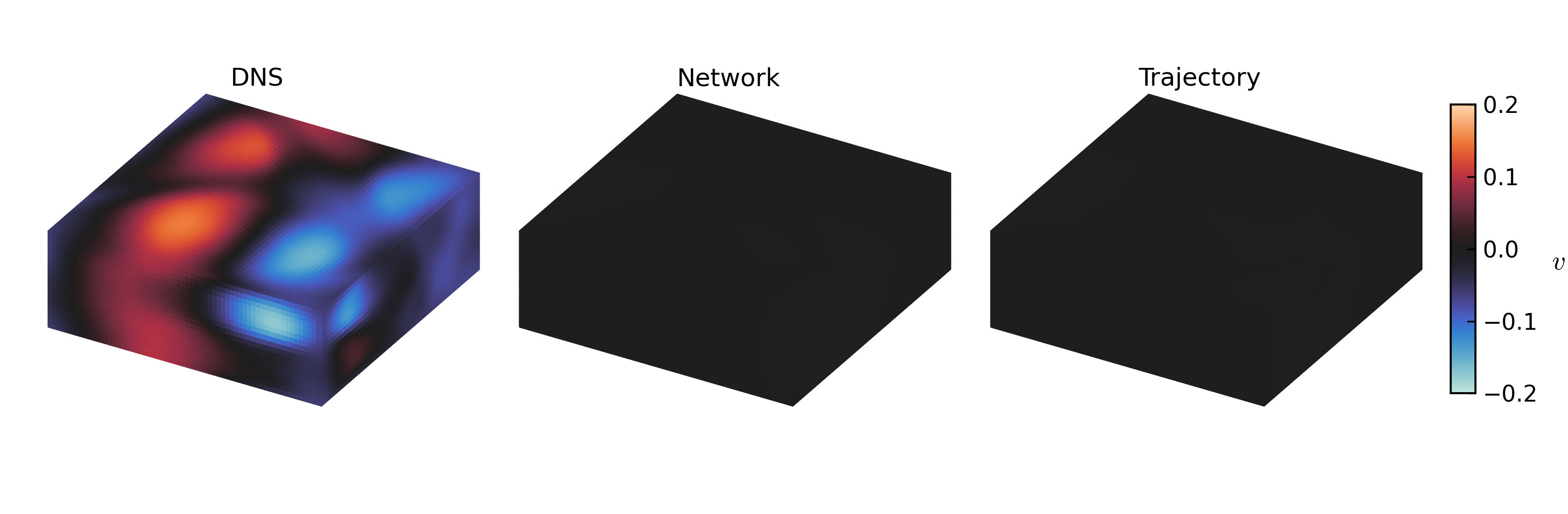}
    \caption{Training/validation losses (top) and output fields (bottom two rows) of a converged PINN applied to the shadowgraph training set described in the main body of the paper. 
    Shown are predicted density (centre) at both initial time, alongside the time-marched predictions advanced by TIME and the predicted vertical velocity (bottom). Note the dramatic under prediction in both quantities; the vertical velocity predicted by the PINN cannot be seen on the color scale here.  
    }
    \label{fig:pinn_fail}
\end{figure}
In figure \ref{fig:pinn_fail} we show the output of a `converged' 11 layer PINN (see training description above) in the shadowgraph case. 
The density field is dramatically underpredicted, with notable qualitative differences in the structures at $t=0$, which persist under time marching. 

\begin{figure}
    \centering
    \includegraphics[width=0.5\linewidth]{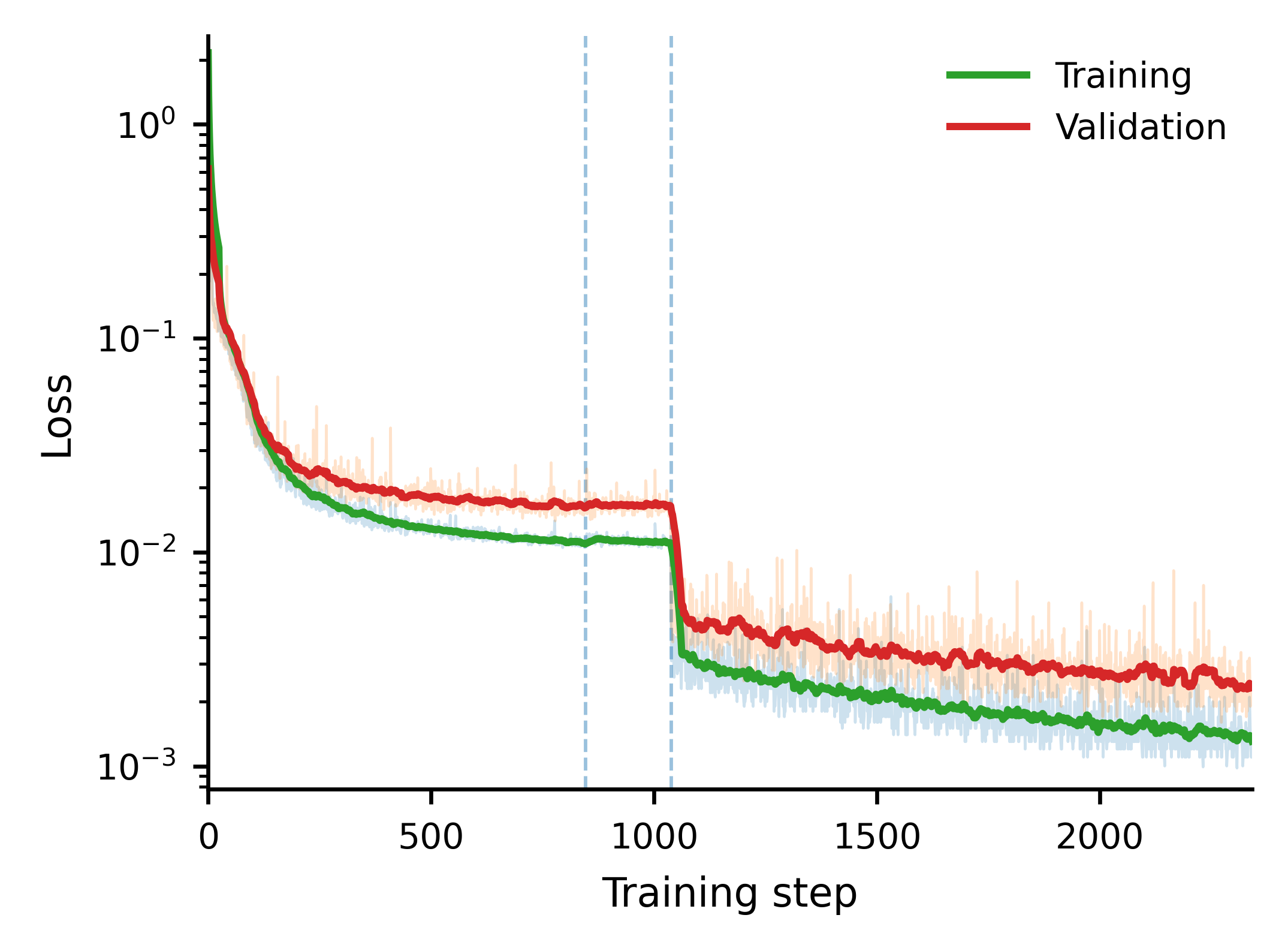}
    \includegraphics[width=\linewidth]{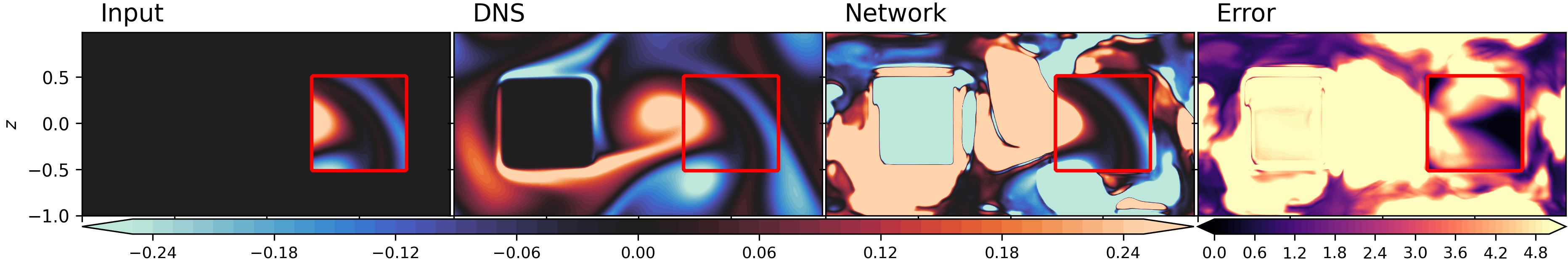}
    \caption{Failure of the pure `assimilation only' loss (equation 1 in the main text) in the partial-to-full field reconstruction.
    Shown are the loss during training (top) and fields (bottom): masked input ($l=1/2$), the ground truth, network output and error. No reasonable solution was found for this algorithm. 
    }
    \label{fig:mask_fail}
\end{figure}
We also present results of the pure `assimilation' training algorithm for the partial-to-full field problem in figure \ref{fig:mask_fail}. 
In this there is dramatic over prediction in the vorticity field everywhere in the domain. 
Resolution of this issue would likely require pre-training of the network offline.
The TraCTra algorithm, on the other hand, is able to locate a realistic solution by virtue of the regularising effect of its self-consistency term (see equation 2 in the main paper).

\bibliographystyle{unsrtnat}
\bibliography{sn-bibliography}

\end{document}